\documentclass[reprint,superscriptaddress,amsmath,amssymb,aps,prb,floatfix]{revtex4-2}
\usepackage{siunitx}
\usepackage{makecell}
\usepackage[colorlinks,linkcolor=black,anchorcolor=blue, urlcolor=blue,citecolor=blue,unicode]{hyperref}
\usepackage{multirow}
\usepackage{subfigure}
\usepackage{color}
\usepackage[normalem]{ulem}
\usepackage{amsmath}
\usepackage{booktabs,siunitx,array}
\usepackage{booktabs}
\usepackage{ulem}
\usepackage{listings}
\usepackage{import}
\usepackage{xcolor}
\usepackage{framed}
\usepackage{mdwlist}
\usepackage{diagbox}
\usepackage{bm}
\usepackage{graphicx}
\usepackage{dcolumn}
\usepackage{algorithm}
\usepackage{algpseudocode}
\usepackage{csquotes}
\usepackage{array}
\usepackage{subfigure}
\usepackage{siunitx}
\usepackage{amsmath}
\usepackage{ifthen}
\usepackage{threeparttable}
\usepackage{float}

\usepackage{tabularx}
\usepackage{CJK}
\usepackage{indentfirst}
\allowdisplaybreaks 

\begin{document}

\title{A finite-element Delta-Sternheimer approach for computing accurate all-electron RPA
correlation energies of polyatomic molecules
}
\author{Hao Peng}
\affiliation{Institute of Physics, Chinese Academy of Sciences, Beijing 100190, China}
\affiliation{School of Physical Sciences, University of Chinese Academy of Sciences, Beijing 100049, China}
\author{Haochen Liu}
\affiliation{SKLMS, ICMSEC, NCMIS, Academy of Mathematics and Systems Science, Chinese Academy of Sciences, Beijing 100190, China}
\affiliation{School of Mathematical Sciences,
University of Chinese Academy of Sciences, Beijing 100049, China }
\author{Chuhao Li}
\affiliation{Institute of Physics, Chinese Academy of Sciences, Beijing 100190, China}
\affiliation{School of Physical Sciences, University of Chinese Academy of Sciences, Beijing 100049, China}
\author{Hehu Xie}
\email{hhxie@lsec.cc.ac.cn}
\affiliation{SKLMS, ICMSEC, NCMIS, Academy of Mathematics and Systems Science, Chinese Academy of Sciences, Beijing 100190, China}
\affiliation{School of Mathematical Sciences,
University of Chinese Academy of Sciences, Beijing 100049, China }
\author{Xinguo Ren}
\email{renxg@iphy.ac.cn}
\affiliation{Institute of Physics, Chinese Academy of Sciences, Beijing 100190, China}

\newcommand{\XR}[1]{\color{red}{\bf #1 }}
\newcommand{\XB}[1]{\color{blue}{ #1 }}
\newcommand{\XY}[1]{\color{yellow}{\bf #1 }}

\begin{abstract}

Attaining a reliable complete basis set (CBS) limit remains a significant challenge in ab initio correlated electronic-structure calculations. Building on our previous work for atoms and diatomic molecules, we present a finite-element (FE) Delta Sternheimer approach for numerically accurate random phase approximation (RPA) calculations applicable to general molecules. This approach seamlessly integrates atomic orbital basis sets with FE grids, enabling an arbitrary precision representation of first order wavefunctions. As a result, the density response function and RPA correlation energies can be computed with fully controlled numerical precision. The Delta Sternheimer approach thus provides direct access to RPA correlation energies at the CBS limit, eliminating reliance on conventional extrapolation schemes.

We apply this approach to two problems: The energy hierarchy of 20 water-dimer configurations and the atomization energies of 50 molecules from the G2 set. For the water dimer, we examine the basis set dependence of the isomer energy ordering. For the G2 set, we investigate the residual numerical uncertainty in the conventional extrapolated CBS limit, both with and without correction for basis-set superposition error (BSSE).

\end{abstract}

\maketitle

\section{Introduction}
The tremendous success of Kohn-Sham (KS) density functional theory (DFT) has laid a solid foundation for computational studies in quantum chemistry and materials science \cite{hohenberg1964,kohn1965}. However, DFT with conventional approximations suffers from intrinsic accuracy limitations. In recent years, much effort has been devoted to developing electronic structure methods with higher accuracies, aiming to better describe the correlation effects between electrons \cite{Szabo/Ostlund:1989,Marin/Reining/Ceperley:2016,Friesner:2005,Ghosh/etal:2018}. These methods have considerably enriched the toolbox of electronic-structure theory and allow practitioners to choose an approach according to the system size under study and the desired accuracy.

However, most correlated electronic-structure methods exhibit slow convergence with respect to the single-particle basis set.
In  M{\o}ller–Plesset  perturbation theory (MP2) \cite{moller1934}, configuration interaction (CI), and coupled-cluster (CC) theories \cite{nesbet1955,cizek1966,bartlett2007}, the $N$-electron wavefunction is expanded in terms of Slater determinants constructed from single-particle orbitals.
Ideally, a complete set of such determinants is desired to span the exact many-body space, but in practice, the space is always truncated due to the incompleteness of the single-particle basis, leading to basis set errors.
Similarly, the random-phase approximation (RPA) and the $GW$ approximation require the evaluation of the density response function \cite{hedin1965,bohm1953}, which depends on a complete set of single-particle states \cite{Adler1962,Wiser1963}; hence, the basis-set incompleteness directly affects their accuracy as well.
In $GW$, the Green’s function in the Lehmann representation likewise relies on the full single-particle spectrum.

Consequently, reaching the complete-basis-set (CBS) limit is a central challenge for correlated calculations. A standard and convenient strategy is to use correlation-consistent Gaussian basis sets developed by Dunning and co-workers and to apply an extrapolation formula to estimate the CBS limit \cite{dunning1989,peterson2000}. On the numerical atomic orbital (NAO) side, NAO-VCC-$n$Z sets have been constructed to provide correlation-consistent NAOs \cite{zhang2013,Yang/Zhang/Ren:2024}. Although extrapolation with correlation-consistent basis sets often reduces basis errors at a modest cost, truly reliable, parameter-free CBS results remain difficult to obtain.

A fundamental source of this difficulty is the electron–electron cusp: Owing to the Coulomb singularity, the many-electron wavefunction has a Kato cusp when two electrons approach each other \cite{kato1957}.  Explicitly correlated ($R_{12}$/$F_{12}$) approaches incorporate interelectronic-distance factors to remove the cusp and thereby dramatically accelerate basis convergence \cite{kutzelnigg1985,kutzelnigg1991_r12_I,termath1991_r12_II,klopper1991_r12_III,tenno2000,klopper2006_review,kong2012explicitly}; these techniques have been successfully applied to CI, MP2, RPA, CC, and related methods \cite{tenno2000,hattig2012,tenno2004,adler2007,adler2011_localF12,knizia2009,humer2022approaching}. However, $F_{12}$/$R_{12}$ methods require many additional electron–electron integrals and can be costly for large systems.

Real-space approaches provide an alternative way to obtain numerically precise results by representing wavefunctions and related quantities directly on a spatial grid or localized real-space basis.
The commonly used real-space discretization schemes include the finite-difference method (FDM), the finite-element method (FEM), and the multiresolution wavelet (MRA) 
method \cite{Chelikowsky1994,Pask2005_fe,Genovese2008_wavelet}. These methods allow for a systematic reduction of discretization errors by controlling a single parameter, thus enabling the convergence of energies and other physical quantities to arbitrary precision. In recent decades, achieving numerically exact electronic structure calculations based on real-space methods has become an important and active research area \cite{Flores1993_FEM_MP2_raregas,Flores1994_FEM_MP2_elements,kottmann2017coupled,kottmann2017wavelets,kottmann2017mp2wavelet}. In our previous works, we developed a real-space Sternheimer method that yields numerically precise all-electron RPA correlation energies for atoms \cite{peng2023basis} and diatomic molecules \cite{peng2024textit}.

Despite these advances, applying real-space methods to correlated many-electron calculations beyond atoms and dimers remains challenging, primarily due to their expensive computational cost. To address this limitation, hybrid strategies that combine real-space grids with localized atomic-orbital (AO) basis offer a promising pathway toward improved efficiency and scalability. In the full-potential linearized augmented plane-wave (LAPW) basis,  Betzinger et al. decomposed the first-order wave function into components inside and outside the original FLAPW basis space \cite{betzinger2012precise,betzinger2013precise,betzinger2015precise}. Within the basis space, the first-order wave function can still be expanded in terms of the FLAPW basis functions, while outside the basis space, it is obtained by solving for the response of the basis functions on a one-dimensional grid. This approach effectively eliminates the basis set error of the FLAPW method in computing the response functions via an ingenious combination of the real-space approach with a finite basis set.

 In this work, we extend our real-space RPA framework from diatomic molecules \cite{peng2024textit} to general molecules using the FEM. Inspired by the work of Betzinger et al. \cite{betzinger2012precise,betzinger2013precise,betzinger2015precise} and the concept of $ \Delta$-learning in the field of machine learning \cite{ramakrishnan2015big}, we develop an approach that combines real-space FE techniques with a finite AO basis set, significantly accelerating the convergence of the energy with the FE grid. We refer to this approach as the \textit{Delta-Sternheimer method}. With this method, we can obtain numerically converged all-electron RPA atomization energies for general molecules, thus eliminating the previous restriction to atomic and diatomic systems. More importantly, the approach of combining real-space techniques with AO holds tremendous application potential to address the
 basis set error issues in other electronic-structure methods such as $GW$.
 We apply the Delta-Sternheimer method to determine the energy ordering of different configurations of water dimers and the atomization energies of G2 molecules. These are used as reference numbers to benchmark the remaining basis set energies of the conventional sum-over-states (SOS) approach with AO basis sets.
 
 The rest of this work is organized as follows. In Section \ref{sec:method}, we present the FE Delta-Sternheimer method developed in this work.
In Section \ref{sec:implement}, we describe the practical implementation details of the FE Delta-Sternheimer method.
In Section \ref{sec:result and disscutions}, we benchmark the FE Delta-Sternheimer method and apply it to determine the energy hierarchy of the water dimers.
Furthermore, we report accurate RPA atomization energies for 50 representative molecules, providing a reliable numerical benchmark for future studies. Section~\ref{sec:summary} concludes this paper.  Further details are provided in the Appendices. In Appendix~\ref{appendix:fem}, the basic concepts of the finite element method and the adaptive mesh refinement technique are introduced. Additional convergence tests and
the details of the computation of the residual function $D_a$ are provided in Appendices~\ref{appendix:converge-test} and \ref{appendix:Da}, respectively.  
In Appendix~\ref{appendix:convergence_RI_error}, we systematically examine the errors associated with the resolution-of-identity (RI) approximation. Finally, additional computational results are presented in Appendix~\ref{appendix:result-sup}.

\section{NUMERICAL METHODS}
\label{sec:method}

\subsection{Basis set incompleteness error in the RPA method}
\label{sec:bsie}
Derived from the adiabatic connection fluctuation-dissipation theorem \cite{Gunnarsson/Lundqvist:1976,Langreth/Perdew:1977}, the RPA correlation energy is given by \cite{Langreth/Perdew:1977,Gunnarsson/Lundqvist:1976},
\begin{equation}
 E_c^\mathrm{RPA}=\frac{1}{2\pi}\int_0^\infty {{\rm d}\omega}\text{Tr}[(\ln(1-\chi^0(i\omega)v)+\chi^0(i\omega)v].
 \label{eq:EcRPA}
 \end{equation}
 where $\chi^0$ and $v$ represent the non-interacting density response function and the bare Coulomb interaction, respectively. 
 Physically, $\chi^0$ describes the density response of the KS system with respect to a change of
 the KS effective potential $v_\text{eff}(\mathbf{r})$. 
 In practical implementation, it is customary to expand the response function $\chi^0(i\omega)$ in terms of an auxiliary basis set (ABS),
\begin{equation}
\chi^0(\mathbf{r},\mathbf{r}',i\omega)=\sum_{\mu,\nu} P_{\mu}(\mathbf{r}) \chi^0_{\mu\nu}(i\omega) P_{\nu}(\mathbf{r}') \, ,
\label{eq:chi0_expan}
\end{equation}
 where $\chi^0_{\mu\nu}(i\omega)$ is the matrix representation of the response function under the ABS.
 Further define the Coulomb matrix as
 \begin{equation}
   V_{\mu\nu} = \iint d\mathbf{r} d\mathbf{r}' \frac{P_{\mu}(\mathbf{r})P_{\nu}(\mathbf{r}')}{|\mathbf{r}-\mathbf{r}'|}\, ,
   \label{eq:coulomb_matrix}
 \end{equation}
 Eq.~\ref{eq:EcRPA} can then be interpreted as a matrix equation with the operators $\chi^0$ and $v$ replaced by their corresponding matrix
 representations under the AFS, as given by Eqs.~\ref{eq:EcRPA} and \ref{eq:coulomb_matrix}. This is the essence of the so-called
 resolution-of-identity (RI) RPA approach \cite{ren2012resolution}.
In practical RI-RPA calculations, the response function is often expanded in terms of the Adler–Wiser  \cite{Adler1962,Wiser1963} formula
 \begin{align}
\chi^0(\mathbf{r},\mathbf{r}',\mathrm{i}\omega)
&= \langle \mathbf{r} | \chi^0(\mathrm{i}\omega) | \mathbf{r}' \rangle \notag \\
&=\sum_{i}^{\mathrm{occ}} \sum_{a}^{\mathrm{unocc}}
\frac{\psi_{i}^*(\mathbf{r}) \psi_{a}(\mathbf{r})
\psi_{a}^*(\mathbf{r}') \psi_{i}(\mathbf{r}')}
{\epsilon_{i} - \epsilon_{a} - \mathrm{i}\omega }
+ \mathrm{c.c.}\, ,
\label{eq:chi_zero}
\end{align}
where the summation over unoccupied KS states in Eq.~\ref{eq:chi_zero} is limited by the finite size of the AO basis set.

Therefore, in the usual RI-RPA calculations, there are two sources of basis set incompleteness errors (BSIEs): The first comes from a limited number of virtual states included in the summation 
in Eq.~\ref{eq:chi_zero}, due to the finite AO basis set used in the calculations. We call this the single-particle basis error (SPBE).
The second comes from the discretization of the response function in terms of the ABS, as given by Eq.~\ref{eq:chi0_expan}. This is called the ABS error or the RI error. 
In Ref.~\cite{peng2023basis}, we rigorously separated these two sources of BSIEs and found that the SPBE is dominant in a typical AO-based RI-RPA calculation, whereas the RI error is only secondary.

For atoms and diatomic molecules, our previous work \cite{peng2023basis,peng2024textit} demonstrated how to eliminate both types of BSIEs. In brief, for the SPBE, instead of using the SOS formula as given by Eq.~\ref{eq:chi_zero}, we compute the density response function following its original definition: The change of the density upon the variation of the effective potential,
\begin{equation}
\chi^0(\bm{r},\bm{r^\prime},i\omega)=\frac{\delta n(\bm{r},i\omega)}{\delta v_{eff}(\bm{r^\prime},i\omega)}, \label{eq:define chi}
\end{equation}
where the density change is determined by the first-order variation of the occupied KS orbitals,
\begin{equation}
    \delta n(\bm{r},i\omega) = \sum_{i}^\text{occ}\psi_i(\bm{r}) \psi_i^{(1)}(\bm{r},i\omega)\, .
\end{equation}
We note that if one expands the first-order wavefunction  $\psi_i^{(1)}(\bm{r},i\omega)$ in terms of a limited number of zeroth-order KS orbitals, the common SOS expression for $\chi^0$ is recovered. However, in the Sternheimer-RPA approach, $\psi_i^{(1)}(\bm{r},i\omega)$ is determined by solving a Sternheimer equation on a dense real-space grid,
\begin{equation}
			      (H^{(0)}-\epsilon_i+i\omega)\psi_{i}^{(1)}(\bm{r},i\omega)=(\epsilon_i^{(1)}-V^{(1)})\psi_i(\bm{r}),
			      \label{eq:st}
		      \end{equation}
where $H^{(0)} $ is the unperturbed KS Hamiltonian and $V^{(1)}$ the perturbation, 
$\psi_i(\bm{r})$ and $\epsilon_i$ are the KS orbitals and orbital energies, and  
$\psi_{i}^{(1)}(\bm{r},i\omega)$ and $\epsilon_i^{(1)}$ are their first-order variations, respectively.    
By progressively refining the real-space grid, the numerical precision of $\psi_{i}^{(1)}(\bm{r},i\omega)$, and hence the density response function and the RPA correlation energy, can be systematically improved. 

In the  Sternheimer-RPA approach, a second aspect is how to model the ``external perturbation", i.e., the variation of the effective potential
$\delta v_{eff}$ in Eq.~\ref{eq:define chi}, or $V^{(1)}$ in Eq.~\ref{eq:st}. In principle, a set of independent functions is needed to capture an arbitrary variation of the effective KS potential. A convenient choice for such independent functions is the ABS, or the RI functions introduced 
in Eq.~\ref{eq:chi0_expan}. Here, the RI functions, on the one hand, serve as the ``external perturbation", and on the other hand, expand the variation of the density. In this formulation, the Sternheimer-RPA approach reduces to an RI-RPA scheme that has only RI error due to the finite size of ABS, but not the SPBE. Since the RI error is marginal compared to the SPBE, such a Sternheimer-RI-RPA approach can already provide highly precise RPA results. Nevertheless, if one wishes to remove the RI error as well, one effective way is to determine the eigen-spectrum of
the combined $\chi^0v$ operator directly via an iterative procedure. In this way, the RI error is fully under control and can be made arbitrarily small,
as demonstrated for atoms and diatomic molecules \cite{peng2023basis,peng2024textit}. 

Although the RI error can be fully eliminated by iteratively diagonalizing the  $\chi^0v$ operator, this approach is extremely expensive. In the present work, we adopt the Sternheimer-RI-RPA approach and focus on eliminating the SPBS in the RPA calculations for general molecules. In this case, we monitor the reduction of RI error by systematically increasing the size of ABS, ensuring that the RI error is negligibly small by employing an extremely large RI basis set. Details will be presented in the Results section.

In our previous works, by exploiting symmetries, the three-dimensional Sternheimer equation was reduced to one-dimensional (atoms) \cite{peng2023basis} and two-dimensional (diatomic molecules) \cite{peng2024textit} ones. 
These equations were then discretized on real-space grids using the FDM, leading to sparse linear systems that are numerically tractable and 
yield results that are arbitrarily accurate.
In the present work, we aim to extend the Sternheimer method to general molecular systems. This requires solving the Sternheimer equation in full three-dimensional real space, which is numerically far more challenging than the one- and two-dimensional cases due to the curse of dimensionality \cite{Bellman1957,Bungartz2004,Chelikowsky1994}. 
To efficiently treat three-dimensional differential equations, two key challenges must be addressed. First, the three-dimensional real space must be discretized effectively, converting the continuous differential equations into a system of linear equations. Second, the dimensionality of the resulting linear system must be reduced as much as possible to mitigate the curse of dimensionality. These aspects will be discussed in the following subsections.

\subsection{The finite element method}
We first clarify why the FDM employed in our previous work \cite{peng2024textit} is not suitable for discretizing three-dimensional real space, particularly for the all-electron calculations. For diatomic molecules, the use of prolate spheroidal coordinates exploits molecular symmetry, allowing a dense distribution of grid points near both nuclei in real space while enabling a transformation to a uniform virtual grid for efficient finite-difference evaluation. This approach simultaneously satisfies the requirements of all-electron accuracy and efficient differentiation.

For general polyatomic molecular systems, however, the spatial singularities associated with the nuclei are irregularly distributed and lack any symmetry. As a result, no efficient three-dimensional grids can be constructed that both resolve the nuclear singularities and remain suitable for finite-difference operations. We therefore abandon the FDM in the present work and instead seek a real-space discretization method better suited for three-dimensional molecular systems.

In this context, the FEM provides an attractive solution. The FEM divides real space into a set of small elements and defines local basis functions—FE shape functions—within each element. 
Any continuous function can then be 
approximated as a linear combination of these local basis functions with any precision.

In the present work, real space is partitioned into tetrahedral elements, and quartic (fourth-order) Lagrange polynomial basis functions are used within each element \cite{Ciarlet1978}. The fundamental principles of FEM are summarized in the Appendix~\ref{appendix:fem}. Once the basis function type and order are fixed, the spatial resolution of the FEM representation depends solely on the mesh density. By refining the mesh into smaller elements, the discretization error can be systematically reduced.

To achieve efficient refinement, the FEM naturally supports adaptive mesh refinement (AMR), which concentrates computational effort in regions with the largest estimated errors. In practice, \textit{a posteriori} error estimators are evaluated for each element, and the mesh is selectively refined in high-error regions—typically near atomic nuclei—thereby enabling accurate all-electron calculations. Further details of the AMR procedure are provided in Appendix~\ref{appendix:fem}.

In summary, the FEM offers the following two key advantages for molecular all-electron calculations:

\begin{enumerate}
 \item Adaptive resolution: AMR provides higher spatial resolution near nuclei, ensuring accurate all-electron precision.

  \item Analytic derivatives: FEM basis functions are analytic polynomials, allowing for exact evaluation of derivatives and accurate treatment of the kinetic-energy operator.
\end{enumerate}


Next, let's discuss in detail how Eq.~\ref{eq:st} is solved within the FE framework.
To solve this three-dimensional differential equation, we place the target molecule at the center of a cubic region. The selection of the cubic region size (i.e., the approximation to infinity) was discussed in Appendix~\ref{appendix:converge-test}. The cubic region is initially discretized into a set of tetrahedra, followed by several times of uniform refinement to produce the initial mesh. Then, employing the AMR technique, we refine the mesh near the atomic nuclei to accurately capture the rapid oscillations of the wavefunctions in these regions. 
Once a sufficiently dense mesh is generated, the discretization of real space is considered complete.

The next step is to discretize the first-order wavefunction. As mentioned earlier, local basis functions are defined on the degrees of freedom associated with the nodes, edges, and faces of each element. Denoting the set of FE basis functions as $\{\phi_k\}$, the first-order wavefunction can then be expressed as a linear combination of these basis functions:
\begin{equation}
\psi_i^{(1)}(\bm{r})=\sum_k^{N_f}u_{ik}^{(1)}\phi_k(\bm{r}).
\label{eq:expand_1}
\end{equation}
The ultimate goal of the FEM is to convert the differential equations in continuous space into a system of linear equations and to obtain the combination coefficients by solving the linear problem.
To achieve this, we substitute Eq.~\ref{eq:expand_1} into Eq.~\ref{eq:st}, then left multiply both sides by $\phi_{k^\prime}$ and perform integration, arriving at:
\begin{equation}
\sum_k^{N_f} [H^{(0)}_{k^\prime k}-(\epsilon_i-i\omega)S_{k^\prime k}]u_{ik}^{(1)}
=\epsilon_i^{(1)}\langle\phi_{k^\prime}|\psi_i\rangle-V_{k^\prime i}^{(1)}.
\label{eq:st_FEM}
\end{equation}
Here, $H^{(0)}_{k^\prime k}=\langle\phi_{k^\prime}|H^{(0)}|\phi_k\rangle$, 
$S_{k^\prime k}=\langle\phi_{k^\prime}|\phi_k\rangle$, 
$V^{(1)}_{k^\prime i}=\langle\phi_{k^\prime}|V^{(1)}|\psi_i\rangle$. 
Eq.~\ref{eq:st_FEM}  essentially represents an expansion of the Sternheimer equation in terms of a  general set of functions, which is not restricted to FE basis functions.  However, due to the highly localized nature of the FE basis functions, the Hamiltonian and overlap matrices ($H$ and $S$) are both highly sparse.
Eq.~\ref{eq:st_FEM} is a linear system of size $N_f×N_f$. Solving Eq.~\ref{eq:st_FEM} yields the coefficient vector $\{ u_{ik}^{(1)}\}$, which in turn allows us to obtain the first-order wavefunction via Eq.~\ref{eq:expand_1}. The accuracy of the first-order wavefunction obtained in this way is determined solely by the density of the  elements. Therefore, by progressively performing adaptive mesh refinement, a converged first-order wavefunction for the system can be attained. This approach effectively avoids the single-particle basis set error introduced by expanding the first-order wavefunction in a finite AO basis set, which cannot be systematically improved to arbitrary precision.

\subsection{The Delta-Sternheimer Method}
In principle, arbitrarily accurate first-order wavefunctions can be obtained by progressively refining the real-space mesh through adaptive refinement. However, the challenge here is to keep the dimensionality of the resulting linear problem as small as possible in order to avoid the curse of dimensionality.

Table~\ref{tab:grid_converge} presents benchmark calculations for methane, silane, propyne, and benzene. 
For methane, approximately 180000 grid degrees of freedom are sufficient to achieve meV-level convergence of the RPA correlation energy. In contrast, for benzene, even with 450000
 degrees of freedom, one can only achieve an accuracy of about 10 meV. These results demonstrate that the dimensionality of the linear problem increases rapidly with the number of atoms.
\begin{table*}[htbp]
\centering
\footnotesize
\setlength{\tabcolsep}{10pt}
\begin{tabular}{ccccc}
\hline
\textbf{Molecule} & \textbf{Grid Level} & \textbf{NOF} & \textbf{S-S} & \textbf{D-S} \\
\hline
\multirow{4}{*}{CH$_4$} & 1 & 94159 & -13.62162 (25.54) & -13.64063 (6.53) \\
& 2 & 123074 & -13.64253 (4.63) & -13.64591 (1.25) \\
& 3 & 182006 & -13.64618 (0.98) & -13.64683 (0.33) \\
& 4 & 238577 & -13.64714 (0.02) & -13.64716 (0.00) \\
\hline
\multirow{4}{*}{SiH$_4$} & 1 & 115407 & -26.59912 (28.37) & -26.61804 (9.45) \\
& 2 & 160219 & -26.62194 (5.55) & -26.62600 (1.49) \\
& 3 & 200047 & -26.62709 (0.40) & -26.62719 (0.30) \\
& 4 & 242887 & -26.62764 (-0.15) & -26.62749 (0.00) \\
\hline
\multirow{4}{*}{C$_3$H$_4$} & 1 & 142178 & -32.64365 (93.83) & -32.72890 (8.58) \\
& 2 & 194536 & -32.73711 (0.37) & -32.73562 (1.86) \\
& 3 & 248803 & -32.73611 (1.37) & -32.73708 (0.40) \\
& 4 & 301035 & -32.73873 (-1.25) & -32.73748 (0.00) \\
\hline
\multirow{6}{*}{C$_6$H$_6$} & 1 & 214459 & -60.33976 (265.78) & -60.62535 (19.81) \\
& 2 & 250419 & -60.36203 (243.51) & -60.62368 (18.14) \\
& 3 & 316203 & -60.48645 (119.09) & -60.60552 (0.02) \\
& 4 & 360963 & -60.55359 (51.95) & -60.60496 (0.58) \\
& 5 & 405935 & -60.60568 (-0.14) & -60.60529 (0.25) \\
& 6 & 458385 & -60.60065 (4.89) & -60.60554 (0.00) \\
\hline
\end{tabular}
\caption{RPA correlation energy convergence with respect to FE mesh density for CH$_4$, SiH$_4$, C$_3$H$_4$, and C$_6$H$_6$. The energy unit is eV. NOF represents the number of degree of freedoms of the FE mesh.  S-S represents the FE Standard-Sternheimer approach, and D-S represents the FE Delta-Sternheimer approach. Values in parentheses denote the absolute deviation (in meV) from the reference energy, which is taken as the D-S result at the highest grid level for each molecule. }
\label{tab:grid_converge}
\end{table*}
To accelerate the convergence of the RPA correlation energy with respect to the FE mesh density, we therefore introduce an improved Sternheimer approach in the present work. This improvement is motivated by the insights gained from our analysis of the characteristics of different basis sets. AO basis sets are highly efficient, using a relatively small number of basis functions to capture most of the Hilbert space, but they lack systematic improvability. FE basis sets, on the other hand, offer systematic convergence but are less efficient. If one can combine the strengths of both basis sets, 
it is possible to  overcome their respective limitations. In particular, if FE basis functions are only used to describe the component of the first-order wave function that is not well captured by the AO basis set, the required mesh density can be significantly reduced. The reason is that the AO basis set already provides a good approximation to the first-order wave function, and hence the residual difference between the exact and AO-approximated first order wavefunctions  $\Delta \psi^{(1)}$ is expected to be spatially smooth and small in magnitude. Representing $\Delta \psi^{(1)}$ in the FE space is much easier than directly representing the full first-order wavefunction, since smoother functions require lower spatial resolution.

Thus, by using the FE basis set solely to represent $\Delta \psi^{(1)}$, we effectively reduce the reliance on a dense FE mesh. We refer to this framework as the Delta-Sternheimer approach. In the following, we provide a detailed derivation of the working equations for the Delta-Sternheimer approach.

To start with, we denote the complete Hilbert space as $\mathcal{H}$.
Consider a chosen  AO basis set that spans a subspace  $\mathcal{H}_\text{AO}$
 of the complete Hilbert space 
 $\mathcal{H}$. The subspace 
$\mathcal{H}_\text{r}$ is defined as the rest of the space that is not covered by $\mathcal{H}_\text{AO}$
 with respect to $\mathcal{H}$ . Solving the ground‐state DFT problem in  $\mathcal{H}_{AO}$ yields the KS orbitals $\left \{ \psi_{p,\text{AO}}  \right \} $ and their energies  $\left \{ \epsilon_{p,\text{AO}}  \right \} $. Here and below, we use $p$ to denote the general molecular KS orbitals, and $i$ and $a$ for the occupied and unoccupied ones, respectively.  We assume that in the complete Hilbert space $\mathcal{H}$, the eigenfunctions and eigenvalues of the DFT Hamiltonian are $\left \{ \psi_{p}  \right \} $ and  $\left \{ \epsilon_{p}  \right \} $.  Furthermore, $\mathcal{H}$ is decomposed into occupied subspaces $\mathcal{H}_\text{occ}$ and unoccupied subspaces  $\mathcal{H}_\text{unocc}$, and analogously$\mathcal{H}_\text{AO}$ into $\mathcal{H}_\text{AO,occ}$ and $\mathcal{H}_\text{AO,unocc}$, respectively. 
 
 Here, we shall make an important assumption: The chosen AO basis set can provide a complete description of the occupied-state space, i.e., $\mathcal{H}_\text{AO,occ} = \mathcal{H}_\text{occ}$. Of course, in practical calculations, a finite AO basis set also introduces slight errors in the description of the occupied states. However, our tests in Appendix~\ref{appendix:converge-test} demonstrate that the occupied-state manifold obtained in this work has negligible influence on the RPA correlation energy. 
 Therefore, in the following derivations, we do not distinguish between $\mathcal{H}_\text{occ}$ and  $\mathcal{H}_\text{AO,occ}$. In other words, we no longer differentiate between  $\left \{ \psi_{i}  \right \} $, $\left \{ \epsilon_{i}  \right \} $ and $\left \{ \psi_{i,\text{AO}}  \right \} $ ,  $\left \{ \epsilon_{i,\text{AO}}  \right \} $.
 Therefore, 
$\mathcal{H}_\text{r}$ is the complementary subspace of  $\mathcal{H}_\text{AO,unocc}$ with respect to $\mathcal{H}_\text{unocc}$.

Next, we  decompose the first-order wavefunction into $\psi_{i,\text{in}}^{(1)}(\bm{r},i\omega)$ -- the contribution from the subspace  $\mathcal{H}_\text{AO}$, and  $\psi_{i,\text{out}}^{(1)}(\bm{r},i\omega)$ -- the contribution from its complementary subspace 
 $\mathcal{H}_\text{r}$, 
\begin{equation}
\psi_{i}^{(1)}(\bm{r},i\omega) = \psi_{i,\text{out}}^{(1)}(\bm{r},i\omega)+ \psi_{i,\text{in}}^{(1)}(\bm{r},i\omega).
\label{eq:1-order-2-parts}
\end{equation}
Then, by inserting Eq.~\ref{eq:1-order-2-parts} into Eq.~\ref{eq:st} and left-multiplying both sides of the equation by an unoccupied orbital  $\psi_{a,\text{AO}}$ in  $\mathcal{H}_\text{AO}$, we obtain
\begin{multline}
\left\langle \psi_{a,\text{AO}} \middle| 
H^{(0)}-\epsilon_{i}+i\omega 
\middle| \psi_{i}^{(1)}(i\omega) \right\rangle \\
= -\left\langle \psi_{a,\text{AO}} \middle| V^{(1)} \middle| \psi_{i} \right\rangle.
\label{eq:H-expand}
\end{multline}
To derive the right-hand side of the equation,  we have used the property that $\left \langle \psi_{i}  | \psi_{a,\text{AO}}  \right \rangle = 0$ since $i$ and $a$ represent the occupied state and unoccupied state, which are orthogonal to each other.
Furthermore, since $\psi_{a,\text{AO}}$ is an eigenfunction of $H_0$ in $\mathcal{H}_{AO}$ (a subspace of $\mathcal{H}$), we can generally write
\begin{equation}
H^{(0)} |\psi_{a,\text{AO}}\rangle = \epsilon_{a,\text{AO}} |\psi_{a,\text{AO}}\rangle + |D_a\rangle \, ,
\label{eq:H0_on_psia}
\end{equation}
where $D_a$ represents the residual function resulting from the diagonalization of $H^{(0)}$ in the finite AO space  $\mathcal{H}_\text{AO}$.  Obviously, $D_a$ lies in the complementary space  $\mathcal{H}_\text{r}$. On the other hand, for the occupied states, we have
\begin{equation}
|D_i \rangle = (H_0 - \varepsilon_{i}) |\psi_{i}\rangle = 0\, ,
\end{equation}
since we have assumed $\mathcal{H}_\text{AO,occ} = \mathcal{H}_\text{occ}$.

By applying $H^{(0)}$ to
$\langle \psi_{a,\text{AO}}|$ (substituting the conjugate of Eq.~\ref{eq:H0_on_psia} into Eq.~\ref{eq:H-expand}), we obtain
\begin{multline}
(\epsilon_{a,\text{AO}}-\epsilon_{i}+i\omega)
\left\langle \psi_{a,\text{AO}} \middle|  \psi_{i}^{(1)}(i\omega) \right\rangle \\
+ \left\langle D_a \middle|  \psi_{i}^{(1)}(i\omega) \right\rangle
= - \left\langle \psi_{a,\text{AO}} \middle| V^{(1)} \middle| \psi_{i} \right\rangle.
\label{eq:St-expand}
\end{multline}


To proceed, it is important to note that the following orthogonality relationship holds,
\begin{eqnarray}
\left \langle D_a |  \psi_{i,\text{in}}^{(1)}(i\omega) \right \rangle & = & 0, \label{eq:ortho1} \\
\left \langle \psi_{a,\text{AO}} |  \psi_{i,\text{out}}^{(1)}(i\omega) \right \rangle & = & 0,  \label{eq:ortho2}
\end{eqnarray}
since $D_a$ lies entirely in the ${H}_\text{r}$ subspace, while $\psi_{a,\text{AO}}$ lies entirely in the ${H}_\text{AO}$ subspace. 
Substituting Eq.~\ref{eq:1-order-2-parts} into Eq.~\ref{eq:St-expand}, and using Eqs.~\ref{eq:ortho1} and \ref{eq:ortho2}, we obtain
\begin{equation}
 \left \langle \psi_{a,\text{AO}} |  \psi_{i,in}^{(1)}(i\omega) \right \rangle =\frac{\left \langle \psi_{a,\text{AO}}|V^{(1)}|\psi_{i} \right \rangle}{\epsilon_{i}-\epsilon_{a,\text{AO}}-i\omega}+\frac{\left \langle D_a |  \psi_{i,out}^{(1)}(i\omega) \right \rangle}{\epsilon_{i}-\epsilon_{a,\text{AO}}-i\omega}.
\end{equation}
Therefore, we arrive at the following expression for $\psi_{i,\text{in}}^{(1)}(i\omega)$ 
in real space,
\begin{multline}
\psi_{i,\text{in}}^{(1)}(\mathbf{r},i\omega) =  
\sum_a^{N_\text{unocc}}\Bigg(
\frac{\langle \psi_{a,\text{AO}}|V^{(1)}|\psi_{i}\rangle}
{\epsilon_{i}-\epsilon_{a,\text{AO}}-i\omega} \\
+ \frac{\langle D_a |  \psi_{i,\text{out}}^{(1)}(i\omega) \rangle}
{\epsilon_{i}-\epsilon_{a,\text{AO}}-i\omega}
\Bigg)\psi_{a,\text{AO}}(\mathbf{r}).
\label{eq:psi_in}
\end{multline}

The first term in Eq.~\ref{eq:psi_in} corresponds to the 
SOS expression of the first-order wavefunction. 
It can be evaluated by diagonalizing $H_0$ within the $\mathcal{H}_{\mathrm{AO}}$, 
yielding the eigenfunction set $\{\psi_{p,\text{AO}}\}$. 
However, the second term reveals something subtle: 
$\psi_{i,\text{in}}^{(1)}$ depends on $\psi_{i,\text{out}}^{(1)}$. $\psi_{i,\text{in}}^{(1)}$ and $\psi_{i,\text{out}}^{(1)}$ are coupled through the residual function $D_a$, indicating that the origin of this coupling lies in the basis set error introduced by diagonalizing $H_0$ within the finite $\mathcal{H}_{\mathrm{AO}}$. 
If the AO space $\mathcal{H}_{\mathrm{AO}}$
were a complete Hilbert space, the residual function $D_a$ would vanish, and consequently the second term would disappear. Therefore, the presence of the second term directly reflects the incompleteness of the AO space.

Combining Eq.~\ref{eq:1-order-2-parts} and Eq.~\ref{eq:psi_in} 
one finally obtains the full expression of
 the first-order wavefunction in the Delta-Sternheimer approach,
\begin{align}
\psi_{i}^{(1)}(\mathbf{r},i\omega) 
=&\ \psi_{i,\text{out}}^{(1)}(\mathbf{r},i\omega) \notag \\
&+ \sum_a^{N_\text{unocc}}
\Bigg(\frac{\left\langle \psi_{a,\text{AO}}\middle|V^{(1)}\middle|\psi_{i}\right\rangle}
{\epsilon_{i}-\epsilon_{a,\text{AO}}-i\omega} \notag \\
&\quad + \frac{\left\langle D_a \middle| \psi_{i,\text{out}}^{(1)}(i\omega) \right\rangle}
{\epsilon_{i}-\epsilon_{a,\text{AO}}-i\omega}
\Bigg)\psi_{a,\text{AO}}(\mathbf{r}).
\label{eq:1-order-total}
\end{align}
Eq.~\ref{eq:1-order-total} indicates that,  once $\mathrm{H}_{\mathrm{AO}}$ 
is given, the first-order wavefunction in the complete space 
is fully determined by the component $\psi_{i,\mathrm{out}}^{(1)}$ 
in the $\mathrm{H}_\text{r}$ subspace. 

Below, we describe briefly how Eq.~\ref{eq:1-order-total} is evaluated in practice. 
First, the FE mesh is refined using AMR techniques.  
Then, the residual function $D_a$ is evaluated at each FE quadrature point. 
The way to evaluate the residual function is discussed in detail 
in Appendix \ref{appendix:Da}. The most important next step is 
to obtain the accurate outside part of the first-order 
wavefunction $\psi_{i,\mathrm{out}}^{(1)}$. 
To this end, we first solve Eq.~\ref{eq:st_FEM} in the FE space to obtain 
the first-order wavefunction $\psi_{i}^{(1),\mathrm{FE-SS}}$ 
using the FE standard Sternheimer method. 
Here, the superscript FE-SS denotes the first-order wavefunction obtained by using the FE standard Sternheimer (SS) method.  After that, one can project $\psi_{i}^{(1),\mathrm{FE-SS}}$ onto the complementary space, yielding $\psi_{i,\mathrm{out}}^{(1),\mathrm{FE-SS}}$.
Substituting $\psi_{i,\mathrm{out}}^{(1),\mathrm{FE-SS}}$ 
into Eq.~\ref{eq:1-order-total} yields the complete first-order 
wavefunction $\psi_{i}^{(1),\mathrm{FE-DS}}$.  
Here, the superscript FE-DS denotes the first-order wavefunction 
obtained by  using the FE Delta-Sternheimer (DS) method according to Eq.~\ref{eq:1-order-total}.

A subtle point that can easily cause confusion is the distinction between the first-order wavefunction obtained in this manner $\psi_{i}^{(1), \mathrm{FE-DS}}$ and the one obtained on the FE grid $\psi_{i}^{(1),\mathrm{FE-SS}}$. The key point is that the AO space is treated exactly in Eq.~\ref{eq:1-order-total}. On a relatively coarse FE grid, the component of $\psi_{i}^{(1),\mathrm{FE-SS}}$ within the AO space may still be inaccurate, while $\psi_{i,\mathrm{out}}^{(1),\mathrm{FE-SS}}$ has already converged, since the outside component of the first-order wavefunction varies smoothly in real space. As a result, the first-order wavefunction obtained from Eq.~\ref{eq:1-order-total} $\psi_{i}^{(1),\mathrm{FE-DS}}$ is already effectively converged, whereas  $\psi_{i}^{(1),\mathrm{FE-SS}}$ itself is not yet sufficiently accurate.  In Section~\ref{sec:convergence_benchmark} below, we will compare the convergence behavior of the standard Sternheimer method and the Delta-Sternheimer method with respect to grid density. It is found that the Delta-Sternheimer method converges faster and more stably with respect to the grid density, as expected.

 
\section{Implementation Details}\label{sec:implement}
In all-electron Sternheimer-RPA calculations, the density response function can generally be represented in two different ways. In the first approach, the RI approximation  is employed, in which a set of RI functions is introduced to expand the density response function. Within this framework, the RI functions themselves can be used as perturbations, and the corresponding first-order density responses are computed. This has been discussed in Section~\ref{sec:bsie}.
In the second approach, one introduces trial perturbations and directly obtains the eigenvalues of the 
$\chi^0v$ operator through iterative diagonalization, thereby avoiding the RI error. Both approaches were discussed in our previous works \cite{peng2023basis,peng2024textit}.

As mentioned in Section~\ref{sec:bsie}, we adopt the RI scheme to represent the density response function in the present work. Given that the dimension of the linear systems considered here reaches the order of several hundred thousand,  iterative diagonalization of the  $\chi^0v$ operator would be extremely expensive. Moreover, Ref.~\citenum{peng2023basis} has demonstrated that the RI errors are significantly smaller than those arising from the single-particle basis.
In Appendix~\ref{appendix:convergence_RI_error}, we explicitly assess the errors associated 
with the RI approximation. Our tests show that these errors 
can be effectively reduced by enlarging the auxiliary basis set and by 
increasing the maximum angular momentum of the RI functions, rendering the overall RI error well under control.

\begin{figure*}[htbp]
   \centering
   \includegraphics[scale=0.50]{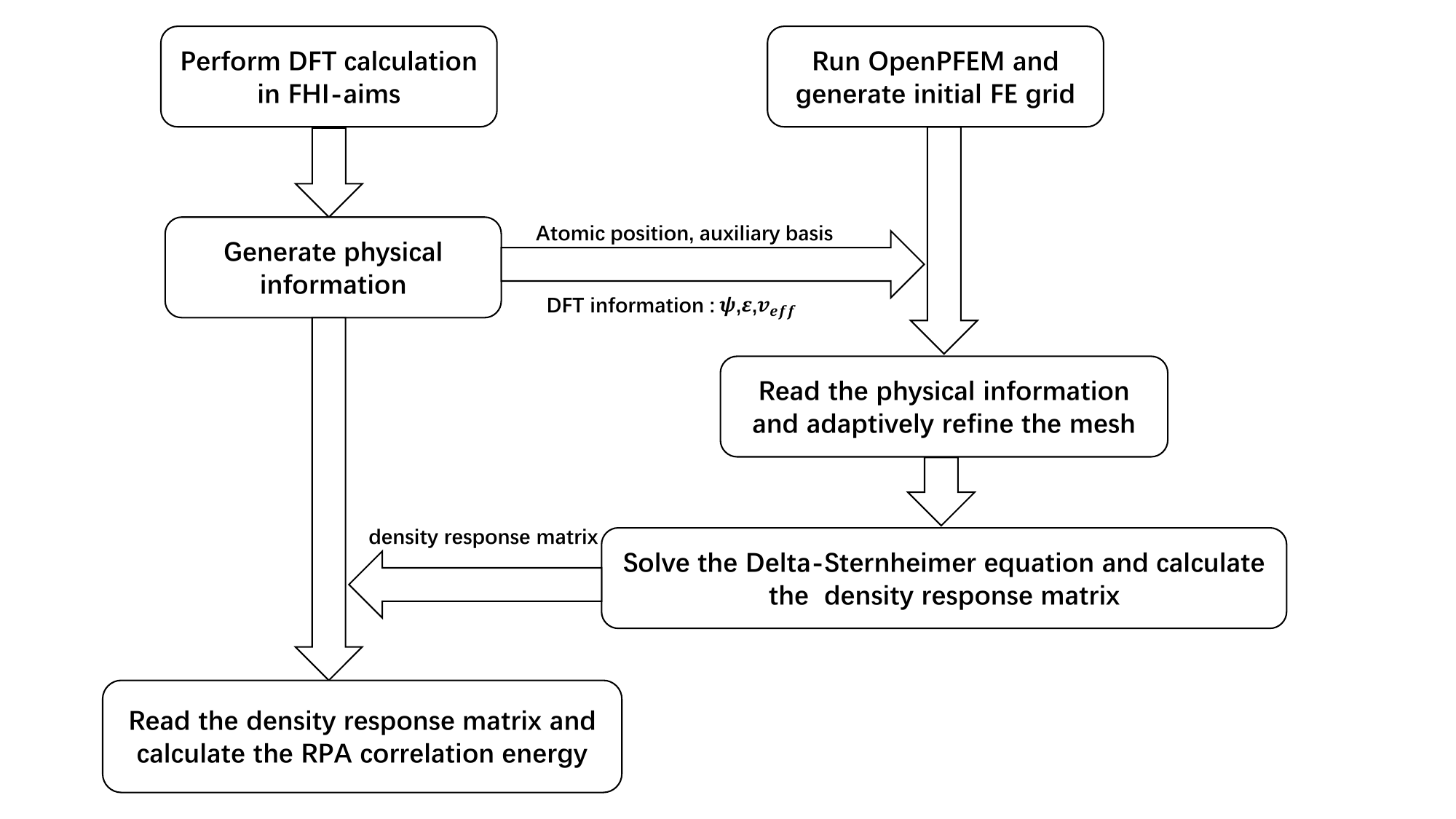}
   \caption{The workflow of actual calculation. FHI-aims primarily provides OpenPFEM with the physical information of the system, including atomic positions, wave functions, eigenvalues, effective potentials, and auxiliary basis sets. After reading these data, OpenPFEM performs AMR and computes the first-order wavefunctions based on Delta-Sternheimer framework. The  density response matrix  is then constructed. FHI-aims subsequently reads the density response matrix  and evaluates the RPA correlation energy using Eq.~\ref{eq:EcRPA}. } 
   \label{fig:workflow}  
\end{figure*}

Next, we briefly describe how the FE Delta-Sternheimer method is implemented, focusing on the actual workflow. We first use FHI-aims \cite{blum2009ab} to perform all-electron DFT \cite{kohn1965self} calculations using the PBE \cite{perdew1996generalized} functional. FHI-aims provides us with all the information for ground-state DFT and the RI basis set. On the FE side, we employ OpenPFEM \cite{Liao2025OpenPFEM}, a parallel FE framework that integrates PETSc \cite{petsc-web-page}, a widely used library for the scalable solution of large-scale sparse linear systems. OpenPFEM facilitates large-scale sparse linear algebra by providing a convenient interface to PETSc-based solvers.  The actual workflow is illustrated in Fig.~\ref{fig:workflow}: First, the ground-state DFT calculation is performed with FHI-aims. Next, the OpenPFEM package is used to generate the FE mesh and carry out AMR. During this process, FHI-aims supplies OpenPFEM with the atomic positions, KS orbitals and eigenvalues, and the KS effective potential. Once the mesh is sufficiently refined, the Sternheimer equation is solved on the dense FE grid, where the RI functions are taken as external perturbations to the system.

After solving the Sternheimer equation, the first-order wavefunction $\psi_{i}^{(1)}$ is projected outside the atomic-orbital (AO) space, yielding the out-of-AO component $\psi_{i,\text{out}}^{(1)}$. The overlap between this component and the precomputed residual function $D_a$ is then evaluated as $\left\langle D_a \middle| \psi_{i,\text{out}}^{(1)} \right\rangle$ .
Finally, by substituting this quantity into Eq.~\ref{eq:1-order-total}, the final first-order wavefunction is obtained. From the resulting first-order wavefunctions, the density-response matrix is assembled by integrating over the FE space. Finally, this density-response matrix is returned to FHI-aims to compute the RPA correlation energy. 

\section{Results and Discussions}
\label{sec:result and disscutions}
In the present work, we intend to obtain highly accurate RPA correlation energies for general molecular systems using the Delta-Sternheimer approach, which is of great scientific interest in two aspects.
First, from a methodological perspective, the proposed approach combines real-space techniques with efficient AO representations, improving computational efficiency without sacrificing accuracy. We believe that such a hybrid strategy, integrating real-space methods with AO basis sets, holds considerable potential for a wide range of electronic-structure methods.
Second, the obtained high-accuracy RPA correlation energies for general molecules provide a reliable reference for assessing finite-basis-set errors. Moreover, these high-quality data can serve as training datasets for the development of accurate machine-learning models.

The rest of this section contains three subsections. In subsection~\ref{sec:convergence_benchmark}, we start by comparing the FE-based standard and Delta-Sternheimer approaches with the two-dimensional prolate spheroidal FDM approach for diatomic molecules developed in our previous work \cite{peng2024textit}. Then, using several molecules as illustrative examples, we present benchmark studies for the convergence behavior of the RPA correlation energy with respect to the mesh density for the standard FE-based Sternheimer scheme and the Delta-Sternheimer approach. This is followed by a further benchmark comparison with the F12 method. In subsection~\ref{sec:water_dimer}, we examine the influence of finite basis set errors on the small energy differences between different configurations of the water dimers. Specifically, we compare the results obtained using the conventional finite-basis SOS method with those obtained in this work using the FE Delta-Sternheimer approach. By testing three widely used correlation-consistent basis sets, we unambiguously assess the impact of finite basis set errors on the RPA energy hierarchy in the water dimers. In subsection~\ref{sec:G2_benchmark}, we select 50 small molecules from the G2/97 set  \cite{curtiss1998gaussian} that contain elements from the first three rows of the periodic table  and compute their RPA atomization energies. These results are again compared with those obtained using the standard SOS approach to assess the finite basis set errors. Meanwhile, we also compute results extrapolated from correlation-consistent basis sets and, by comparing these with our Delta-Sternheimer results, assess the uncertainty in the extrapolation-based CBS limit.

\subsection{Converge behavior with respect to the FE mesh density}
\label{sec:convergence_benchmark}
To validate the numerical accuracy of our method, we first compare the absolute AE RI-RPA correlation energies obtained using the FE–Sternheimer method developed in this work with those obtained using the previously developed basis-error-free RPA approach for diatomic molecules, where the Sternheimer equation is solved in a 2D prolate-spheroidal coordinate system  \cite{peng2024textit}. As a concrete example, we test the ${\rm N}_2$ molecule, using the GTO aug-cc-pwCV5Z for the DFT calculations, with the corresponding auxiliary RI basis generated on-the-fly using the same GTO basis set \cite{ren2012resolution}. For the imaginary frequency integration, 32 modified Gauss–Legendre grid points \cite{ren2012resolution} were employed. Moreover, for the prolate-spheroidal coordinate system, the grid parameters were set to $N_{\mu} = 192$ and $N_{\nu} = 150$, ensuring that the correlation energy is converged to the sub-meV level. This result is therefore taken as the reference for the FE calculations. For the FE calculations,  we computed the RPA correlation energy using both the standard Sternheimer and Delta-Sternheimer approaches with progressively refined FE grids, so that one can verify whether the FE method can converge correctly to the reference value.  The results are presented in Table~\ref{tab:N2_bench}.
\begin{table}[htbp]
\centering
\footnotesize
\begin{tabular}{cccc}
\hline
\textbf{Grid level} & \textbf{NOF} & \textbf{S-S} & \textbf{D-S} \\
\hline
1 & 114139 & -23.10559(106.85) & -23.19675(15.69) \\
2 & 142195 & -23.21279(-0.35) & -23.21094(1.50) \\
3 & 178491 & -23.21273(-0.29) & -23.21149(0.95) \\
4 & 233917 & -23.21168(0.76) & -23.21196(0.48) \\
\hline
\end{tabular}
\caption{The RPA correlation energy of N$_2$ obtained using the FE standard Sternheimer (S-S) and Delta-Sternheimer (D-S) approaches. The reference result ($-23.21244$ eV) is obtained using the 2D Sternheimer approach with prolate spherical coordinate system. The unit in the table is eV. The values in parentheses denote the deviations from the reference result, in units of meV. }
\label{tab:N2_bench}
\end{table}

From Table~\ref{tab:N2_bench}, one can see that, for the present test, the Delta-Sternheimer method converges to the meV level for the absolute all-electron RPA correlation energy already at grid level 3. At grid level 4, the Delta-Sternheimer result differs from the reference value only by 0.48 meV. For the standard Sternheimer method, the error is one order of magnitude larger than that of the Delta-Sternheimer method at grid level 1. However, the standard Sternheimer result also converges to the reference value within 1 meV at grid level 4, although the convergence process somehow shows an unsystematic (zigzag) behavior.
 


\begin{figure}[htbp]
   \centering
   \includegraphics[scale=0.30]{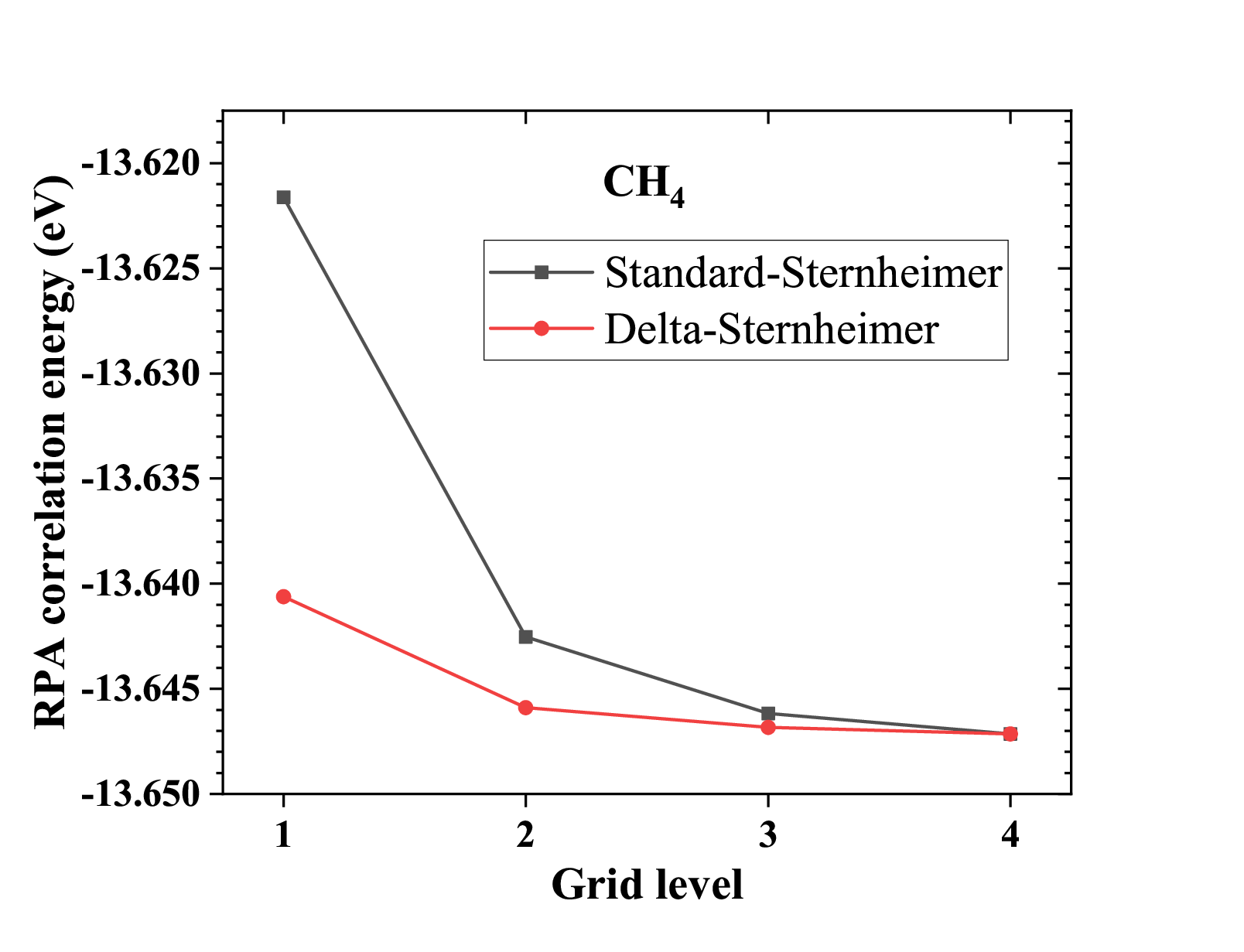}
   \caption{Convergence of the RPA@PBE correlation energy of CH$_4$ with respect to the FE mesh density, obtained using the standard Sternheimer and Delta-Sternheimer approaches. The number of degrees of freedom at each level is given in Table ~\ref{tab:grid_converge} } 
   \label{fig:FEM_grid}  
\end{figure}

Having established the reliability of the FE Sternheimer RPA methods for diatomic molecules, we examine the convergence behavior of the absolute all-electron RPA correlation energies with respect to the FE grid density for four polyatomic molecules -- methane (CH$_4$), silane (SiH$_4$), propyne (C$_3$H$_8$), and benzene (C$_6$H$_6$).
The computational setups here are as follows: In analogy to the N$_2$ case, DFT calculations are performed again using the aug-cc-pwCV5Z basis set, and the auxiliary RI basis set is generated on-the-fly using
the procedure developed in Ref.~\cite{ren2012resolution}. For the integration over the imaginary frequency, 32 Gauss-Legendre grid points are used. The size of the FE box is set to 40 Bohr. 
Figure~\ref{fig:FEM_grid} shows the convergence behavior of the RPA@PBE correlation energy of the methane molecule with respect to the
FE grid density.
It can be readily seen that, at the same grid density, the Delta-Sternheimer method yields lower correlation energies and converges much faster with respect to the grid refinement, compared to the standard Sternheimer method. We also observe that when the grid is relatively coarse—for example, at Grid level = 1, the energy difference between the two methods is rather large --  about 20 meV. This can be understood as the finite AO space used in the Delta-Sternheimer method provides an effective compensation to the incomplete FE space. At Grid level = 4, the correlation energies obtained by the standard Sternheimer and Delta-Sternheimer methods differ by only 0.02 meV, which indicates that the FE space is already very close to completeness and that the finite AO space in the Delta-Sternheimer method is almost fully contained in the FE space. As a result, the two methods yield essentially identical results. 

In Table~\ref{tab:grid_converge}, we report the number of degrees of freedom (NOF, i.e., the total number of basis functions in the FE space) for each grid level in Fig.~\ref{fig:FEM_grid}, along with the RPA correlation energies obtained using both methods. Results for the other three molecules are also presented. 
It can be seen that the results for silane are similar to those for methane. For propyne, however, the standard Sternheimer method shows a certain degree of numerical instability: at Grid levels 2, 3, and 4, the RPA correlation energies fluctuate on the meV scale.
The case of benzene, the largest molecule studied in this work, is of particular importance. First, we found that for the standard Sternheimer method, even with as many as 450,000 grid degrees of freedom, the correlation energy converges only to the 10 meV level and exhibits relatively large numerical fluctuations. This demonstrates that for molecules with a larger number of atoms, the computational cost of using the standard Sternheimer method becomes extremely high. In contrast, the Delta-Sternheimer method already reaches meV-level convergence at Grid level 3.
We further note that, for benzene, the RPA correlation energy obtained by the Delta-Sternheimer method does not decrease monotonically but instead converges upward from a lower value, which differs from the results for the three smaller molecules. Moreover, at Grid levels 3, 4, 5, and 6, small numerical fluctuations below 1 meV are observed. Such small numerical fluctuation
may originate from inherent issues associated with higher-order FE  basis functions \cite{BabuskaSuri1994}. Nevertheless, in the Delta-Sternheimer method, this phenomenon appears only for the relatively large benzene molecule, and the magnitude of the fluctuation remains extremely small. Therefore, it does not affect the conclusion that the Delta-Sternheimer method can significantly accelerate the convergence of the RPA correlation energy with respect to the FE grid density and provides better numerical stability.
Another important conclusion is that, for benzene, the largest molecule studied in this work, the RPA correlation energy obtained with the Delta-Sternheimer method converges to the meV level at a grid density corresponding to about 300,000 degrees of freedom. Based on this test, for all of the small molecules investigated in this work, the FE grid degrees of freedom were adaptively refined to at least 300,000.

Finally, we compare the molecular atomization energies at the RPA level obtained in this work with the high-accuracy results reported by Humer et al. \cite{humer2022approaching} using the F12 method.
Humer et al. systematically investigated the basis-set convergence of the RPA correlation energy by combining plane-wave projector augmented-wave (PAW) calculations with explicitly correlated Gaussian-type orbital (GTO) methods \cite{humer2022approaching}. In the GTO part, the introduction of the F12 approach accelerates the basis-set convergence of the RPA correlation energy, thereby yielding results close to the basis-set limit. In Ref.~\cite{peng2024textit}, we have already verified the high accuracy of the F12 results for diatomic molecules. In the present work, we perform further benchmark comparison for polyatomic molecules. 
Here, we still perform the DFT calculations using the aug-cc-pwCV5Z basis set. The auxiliary basis is generated on the fly, with the maximum angular momentum of the auxiliary functions set to 9. The KS orbitals from the DFT calculations are used not only as input to the Delta-Sternheimer framework but also to 
evaluate the non-self-consistent Hartree–Fock (HF) part of the RPA total energy.

Subsequently, the Delta-Sternheimer equations are solved following the workflow displayed in Fig.~\ref{fig:workflow}, and the RPA correlation energy is evaluated. The final total atomization energy is obtained as the sum of the RPA correlation contribution and the non-self-consistent Hartree–Fock contribution.
Here, the NOF of the FE mesh is set to 300,000 according to the convergence test of section\ref{sec:convergence_benchmark}.  All parameter settings are the same as in Sec.~\ref{sec:G2_benchmark}.

\newcolumntype{Y}{>{\centering\arraybackslash}X} 
\newcolumntype{C}{>{\centering\arraybackslash}X} 

\begin{table*}[htbp]
\centering
\scriptsize
\renewcommand{\arraystretch}{1.2}
\begin{tabularx}{\textwidth}{|Y|C|C|C|C|C|C|}
\hline
Molecule & F12 AE & This work AE & $\Delta$(AE) & F12 FC & This work FC & $\Delta$(FC) \\
\hline
HCN        & -300.70 & -300.54 & -0.16 & -298.87 & -298.69 & -0.18 \\
H$_2$O     & -223.60 & -223.57 & -0.04 & -223.19 & -223.26 &  0.07 \\
H$_2$O$_2$ & -256.34 & -256.30 & -0.04 & -255.81 & -255.78 & -0.03 \\
CH$_4$     & -405.94 & -406.10 &  0.16 & -494.77 & -494.93 &  0.16 \\
C$_2$H$_6$ & -685.92 & -686.11 &  0.19 & -683.68 & -683.89 &  0.21 \\
C$_2$H$_2$ & -383.62 & -383.47 & -0.15 & -381.17 & -381.02 & -0.15 \\
C$_2$H$_4$ & -539.60 & -539.70 &  0.10 & -537.30 & -537.41 &  0.11 \\
CO$_2$     & -366.85 & -366.44 & -0.41 & -364.76 & -364.51 & -0.25 \\
CH$_4$O    & -492.23 & -492.28 &  0.05 & -490.83 & -490.92 &  0.09 \\
CH$_2$O    & -357.08 & -356.97 & -0.11 & -355.69 & -355.62 & -0.07 \\
NH$_3$     & -291.25 & -291.30 &  0.05 & -290.55 & -290.65 &  0.10 \\
N$_2$H$_4$ & -427.82 & -427.91 &  0.09 & -426.63 & -426.72 &  0.09 \\
\hline

MD         &   -     &    -    & -0.02 &     -    &    -     &  0.01 \\
MAD        &   -     &    -    &  0.13 &     -    &    -     &  0.13 \\
\hline
\end{tabularx}
\caption{RPA atomization energies obtained using the F12 method (Ref.~\cite{Humer/etal:2022}) and the Delta-Sternheimer method (this work). Both the all-electron (AE) and frozen-core (FC) results are presented. MD and MAD indicate the mean deviation (MD) and mean absolute deviation (MAD) between the two methods. The energy unit is  kcal/mol, rounded to two decimals. }
\label{tab:F12-converted}
\end{table*}

Table~\ref{tab:F12-converted} shows that the atomization energies obtained using the F12 method \cite{Humer/etal:2022} and those obtained using the FE Delta-Sternheimer method are fairly close for most of the molecules. The mean absolute deviation (MAD) for this set of molecules is only 0.13 kcal/mol (about 6 meV), verifying that the F12 method can indeed effectively eliminate the finite basis set errors.
In Table~\ref{tab:F12-converted}, both the all-electron and frozen-core results are presented. We would like to mention that the frozen-core approximation is a widely used protocol in practical RPA calculations. However, our previous work \cite{peng2024textit} and Table~\ref{tab:F12-converted} show that this approximation can incur an error of over 1 kcal/mol in the RPA atomization energy of small molecules.

\subsection{Influence of basis set error on the energy hierarchy of water dimers }
\label{sec:water_dimer}
In this subsection, we apply the FE Delta-Sternheimer approach to calculate the RPA energy calculations for a set of water dimers, for which the energy differences between different configurations are exceedingly small. Our primary objective is to quantify how well these subtle energy differences are reproduced by different single-particle basis sets, and through this benchmark, to assess the magnitude of the basis set incompleteness error. Although RPA has been shown to excel in discerning low-energy isomers of small water clusters \cite{Chedid/Jocelyn/Eshuis:2021,Tahir/etal:2022,Tahir/etal:2024,Tahir/etal:2025}, the extent to which the energy hierarchy predicted by RPA is affected by single-particle basis sets remains largely unexplored.  To this end, we randomly selected 20 configurations from a 200-step molecular-dynamics trajectory of the water dimer. For each configuration, we perform adaptive mesh refinement in the FE representation to obtain  numerically converged  AE RPA correlation energies. The RPA total energies are obtained by adding the HF energy (evaluated with Kohn-Sham orbitals), 
computed using the highly converged FHI-aims-2010 (\textbf{\textit{tier}}) NAO basis sets \cite{blum2009ab}.
The total RPA energies obtained in this way are used as reference values to benchmark those calculated using the usual SOS approach with finite atomic basis sets. Here, results to be compared were calculated using the SOS approach with three families of correlation-consistent GTO/NAO basis sets: (1) aug-cc-pwCVXZ, (2) cc-pVXZ, and (3) NAO-VCC-$n$Z \cite{dunning1989gaussian,peterson2002accurate,peterson2002systematically,zhang2013numeric}.
Furthermore, we note that all calculations in this subsection (both the Delta-Sternheimer and SOS ones) were performed using the same auxiliary (RI) basis sets.
Specifically, we employed the largest RI-fitting basis sets available from the Basis Set Exchange website: aug-cc-pwCV5Z-RIFIT for oxygen and aug-cc-pV6Z-RIFIT for hydrogen\cite{pritchard2019basis,dunning1989gaussian,kendall1992electron,peterson2002accurate,peterson2002systematically}. 
The choice of identical RI basis sets across all calculations is intentional, as our primary focus is on the error introduced by the single-particle basis sets, not the RI approximation. Therefore, the use of the largest available RI-fitting basis sets ensures that the RI error is negligible.

\begin{figure*}[htbp]
   \centering
    \includegraphics[scale=0.4]{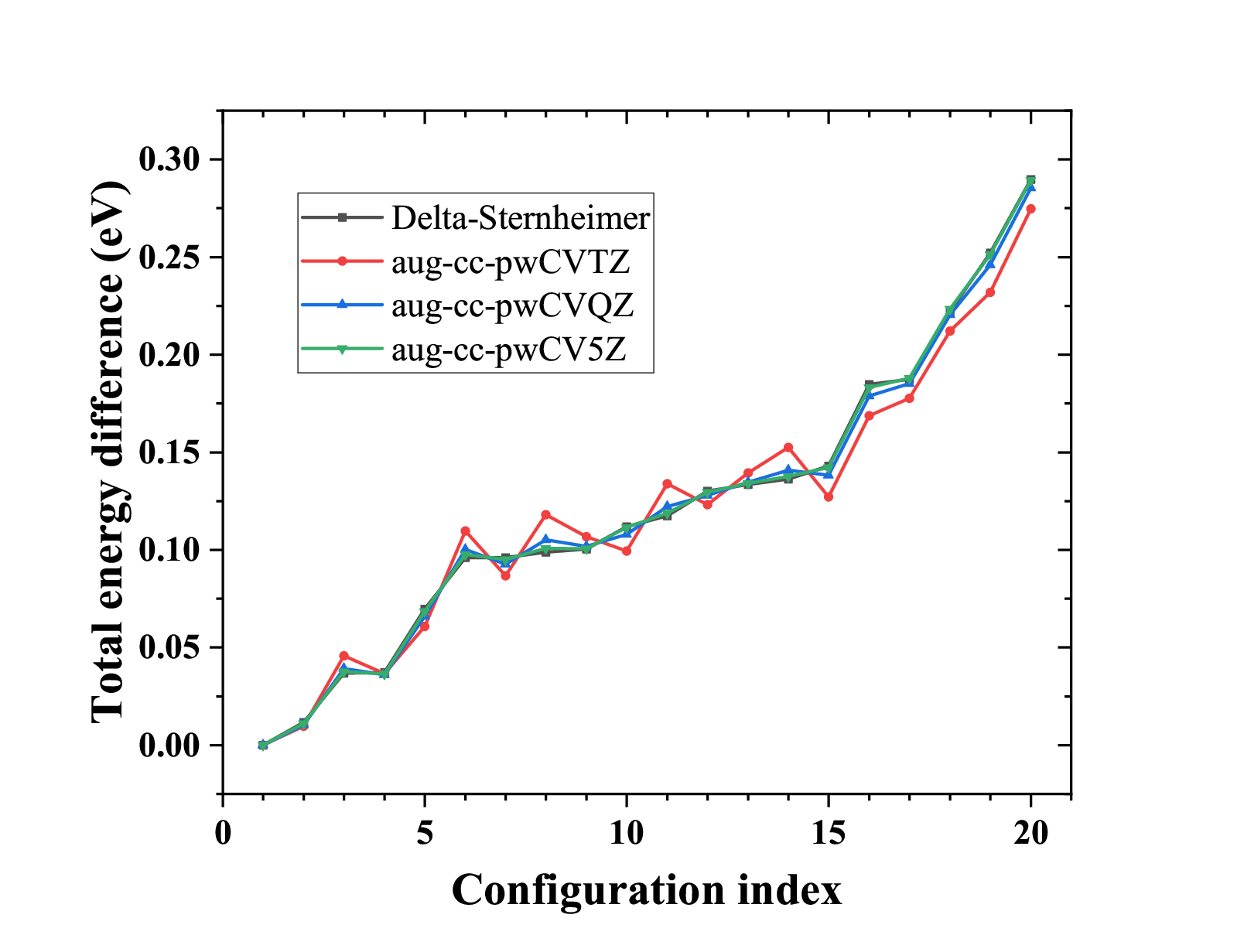}
    \caption{ The RPA total energy differences of 20 different configurations of the water dimer. 20 different configurations of the water dimer are ordered from lowest to highest according to their RPA total energies calculated in the present work. The black line corresponds to the results of this work and is therefore monotonically increasing by construction. The other three lines represent SOS results obtained with the aug-cc-pwCVXZ series of basis sets. }
    \label{fig:water-dimer}
\end{figure*}

In Fig.~\ref{fig:water-dimer}, We rank the 20 configurations by the RPA total energy computed by the FE Delta–Sternheimer method, from low to high, where the energy of the lowest-energy configuration is set to zero, and the energy differences of the other configurations relative to this reference are plotted. We then compared the results obtained with the SOS method based on different correlation-consistent basis sets to those obtained using the Delta-Sternheimer method. In Fig.~\ref{fig:water-dimer}, only the results for aug-cc-pwCVXZ (X=T,Q,5) are shown, whereas the results obtained using the two other series of basis sets are presented in Fig.~\ref{fig:water-dimer-sup} in Appendix~\ref{appendix:result-sup}.
From Figs.~\ref{fig:water-dimer} and \ref{fig:water-dimer-sup}, it can be seen that as the basis set size increases, the energy difference curves obtained from aug-cc-pwCVXZ and NAO-VCC-$n$Z basis sets gradually approach the reference results produced by the FE Delta-Sternheimer method. These figures also reveal that the aug-cc-pwCVTZ (and NAO-VCC-3Z, see Appendix~\ref{appendix:result-sup}) basis set still yields an incorrect total-energy ordering for some configurations; this problem essentially disappears with aug-cc-pwCVQZ and aug-cc-pwCV5Z (NAO-VCC-4Z and NAO-VCC-5Z). However, for cc-pVXZ (X=Q,5,6) series of basis sets, even with the largest cc-pV6Z, the energy
orderings of some of the configurations are still not entirely correct, with errors in the order of meV.
This suggests the importance of diffuse Gaussian functions for accurately describing the energetics of water clusters, which are missed in cc-pVXZ.

To quantify the basis set error, we evaluate the MAD from the reference energy over 19 configurations as follows
\begin{equation}
    {\Delta E}_\text{MAD}=\sum_{j=2}^{20}\frac{\left | {\Delta E}_{j}-  {\Delta E}_{j}^\text{ref}\right |}{19} \, ,
\end{equation}
where $\Delta_j = E_j-E_1 $, and $E_j$ and $E_1$ represent the energies of the $j$-th configuration and the lowest-energy configuration, respectively. The superscript ``ref" denotes the energy obtained from the FE Delta-Sternheimer method, while the values without a superscript correspond to the energies calculated using different correlation-consistent basis sets. We can also separately assess the basis set errors for the two components of the RPA total energy, i.e., the HF total energy and the RPA correlation energy. The results are shown in Table~\ref{tab:basis_comp_all}.
\begin{table*}[htbp]
\centering
\footnotesize
\renewcommand{\arraystretch}{1.2}
\begin{tabular}{l rrr rrr rrr}
\toprule
\multirow{2}{*}{\makecell[c]{Energy\\part}} 
  & \multicolumn{3}{c}{aug-pwCVXZ} 
  & \multicolumn{3}{c}{NAO-VCC-nZ} 
  & \multicolumn{3}{c}{cc-pVXZ} \\
\cmidrule(lr){2-4}\cmidrule(lr){5-7}\cmidrule(lr){8-10}
  & TZ & QZ & 5Z & 3Z & 5Z & 5Z & QZ & 5Z & 6Z \\
\midrule
RPA   & 7.01 & 2.69 & 0.78 & 5.85 & 4.30 & 1.16 & 7.49 & 5.45 & 3.95 \\
HF  & 4.24 & 0.76 & 0.23 & 2.36 & 1.28 & 0.78 & 1.97 & 0.64 & 0.66 \\
TOTAL & 11.24 & 3.45 & 0.97 & 4.83 & 3.17 & 1.65 & 7.27 & 5.18 & 4.16 \\
\bottomrule
\end{tabular}
\caption{Comparison of basis set error for different basis sets. The energy unit is meV.}
\label{tab:basis_comp_all}
\end{table*}
 In brief, Table~\ref{tab:basis_comp_all} shows that when all three basis-set families are taken at their largest level, the HF  errors are all below 1 meV. This is because the HF total energy depends primarily on the description of the occupied-state manifold and hence the basis-set incompleteness error is negligible.
The errors are noticeably larger for the RPA correlation part. In particular, the error of cc-pV6Z is still about 4 meV, consistent with its inferior performance in discerning the energy orders of water dimers in Fig.~\ref{fig:water-dimer-sup}.

\subsection{Atomization energies }
\label{sec:G2_benchmark}

In this subsection, we report the RPA atomization energies of 50 molecules selected from the G2/97 set obtained by the FE Delta-Sternheimer method. All molecular geometries were taken from Ref.~\cite{haunschild2012new}. The computational details are described below. The HF part and the RPA correlation part of the total energy are computed separately. For the former, we performed calculations using FHI-aims with the aug-cc-pwCV5Z basis set. The error in the HF total energy mainly originates from the basis-set incompleteness of the atomic orbitals used to describe the occupied manifold. To assess the HF error for atomization energies, we performed calculations using aug-cc-pwCVQZ and aug-cc-pwCV5Z basis sets, and extrapolated the results to the CBS limit using a standard formula.  In addition, calculations with the built-in \textbf{\textit{tier}} basis sets of FHI-aims were also performed as a reference. For further details see Table \ref{tab:exx-bind} in Appendix \ref{appendix:result-sup}.

From Table \ref{tab:exx-bind} in Appendix \ref{appendix:result-sup}, it is evident that the results obtained with the aug-cc-pwCV5Z basis set agree very well with both the extrapolated values and those based on the \textit{tier}4 NAO basis set. The data in the last row indicate that the HF contribution to the atomization energy obtained with the largest GTO basis differs from the extrapolated result by only 0.03 kcal/mol and from the largest NAO result by 0.02 kcal/mol (about 1 meV). This demonstrates that the aug-cc-pwCV5Z basis set provides a description of the occupied states that is already very close to the CBS limit for atomization energies. It also indicates that the choice of different types of atomic orbitals incurs very little uncertainty in the description of the occupied states.  These results are consistent with the water dimer case.

For the RPA correlation energy, we solved the Delta-Sternheimer equation in the FE space to obtain the density-response matrix and the correlation energy. The main errors here can be attributed to three sources: (1) the FE grid density, (2) the RI approximation, and (3) the effect of basis set errors in the occupied orbitals and energies on the RPA correlation energy when solving the Delta-Sternheimer equation. The convergence with respect to the FE grid density has been discussed in subsection \ref{sec:convergence_benchmark}, while the other two aspects are addressed separately in Appendix \ref{appendix:convergence_RI_error} and Appendix \ref{appendix:converge-test}. The results show that the errors from the FE grid density and from the occupied-state orbitals and energies are both below 1 meV and thus negligible compared with the RI error. The RI approximation, therefore, becomes the dominant basis error source in this work. To minimize the RI error, we employed auxiliary bases generated on-the-fly from aug-cc-pwCV5Z with $L_{\text{max}}=9$. Tests on diatomic molecules indicate that the average error for atomization energies is only 0.04 kcal/mol (about 2 meV). Since the RI error is expected to scale linearly with the number of atoms, it can be estimated as about 0.02 kcal/mol (about 1 meV) per atom. Put together, after considering all possible error sources, we conclude that the dominant error in this work arises from the RI approximation, amounting to roughly 1 meV per atom.

In Tab.~\ref{tab:rpa-bind}, we report the total atomization energies of 50 molecules at the RPA level (including both the HF and RPA correlation contributions). These results can serve as reference data for future benchmark studies and provide the numerical standards for the development of correlation-consistent basis sets. In addition, to assess the basis set errors of finite-basis calculations, we present results obtained with the aug-cc-pwCV5Z basis set within the SOS approach. Furthermore, we report CBS-extrapolated results obtained by applying the standard extrapolation formula  to the SOS results from aug-cc-pwCVQZ and aug-cc-pwCV5Z, which allow us to evaluate the accuracy of the extrapolation schemes.
\newcolumntype{Y}{>{\centering\arraybackslash}p{3.2cm}} 
\newcolumntype{C}{>{\centering\arraybackslash}X} 
\begin{table*}[htbp]
\centering
\scriptsize
\renewcommand{\arraystretch}{1.2}
\begin{tabularx}{\textwidth}{|Y|C|C|C|C|C|C|}
\hline
MO & This work AE & This work FC & 5Z AE & 5Z FC & Extrapolation AE & Extrapolation FC \\
\hline
CH$_4$       & -406.10 & -404.93 & -405.31 (0.79) & -404.17 (0.76) & -406.35 (-0.25) & -405.17 (-0.24) \\
C$_2$H$_2$   & -383.47 & -381.02 & -381.99 (1.49) & -379.61 (1.41) & -383.91 (-0.44) & -381.42 (-0.40) \\
C$_2$H$_4$   & -539.70 & -537.41 & -538.26 (1.44) & -536.04 (1.37) & -540.08 (-0.38) & -537.77 (-0.36) \\
C$_3$H$_6$(Propene)& -823.79 & -820.43 & -821.63 (2.16) & -818.30 (2.12) & -824.35 (-0.56) & -820.90 (-0.47) \\
C$_3$H$_6$(Cyclopropane)& -816.32 & -815.37 & -813.92 (2.40) & -810.59 (4.77) & -816.75 (-0.43) & -813.30 (2.07) \\
C$_3$H$_4$(Allene)& -668.66 & -665.09 & -666.58 (2.08) & -663.08 (2.00) & -669.08 (-0.42) & -665.44 (-0.35) \\
C$_3$H$_4$(Propyne)& -669.82 & -666.23 & -667.62 (2.20) & -664.09 (2.14) & -670.32 (-0.50) & -666.64 (-0.41) \\
C$_3$H$_4$(Cyclopropene)& -647.45 & -644.04 & -645.19 (2.26) & -641.85 (2.19) & -647.88 (-0.42) & -644.40 (-0.36) \\
C$_2$H$_6$   & -686.11 & -683.89 & -684.54 (1.57) & -682.38 (1.51) & -686.51 (-0.40) & -684.26 (-0.37) \\
CH$_2$O      & -356.97 & -355.62 & -355.69 (1.27) & -354.36 (1.26) & -357.17 (-0.21) & -355.79 (-0.17) \\
CH$_3$OH     & -492.28 & -490.92 & -490.79 (1.49) & -489.44 (1.48) & -492.57 (-0.29) & -491.18 (-0.26) \\
CH$_2$O$_2$  & -475.22 & -473.42 & -473.18 (2.03) & -471.43 (1.99) & -475.51 (-0.30) & -473.70 (-0.28) \\
C$_2$H$_2$O$_2$ & -600.27 & -597.52 & -597.73 (2.54) & -595.02 (2.50) & -600.63 (-0.36) & -597.81 (-0.30) \\
C$_2$H$_6$O  & -776.84 & -774.42 & -774.68 (2.16) & -772.25 (2.17) & -777.36 (-0.52) & -774.86 (-0.44) \\
C$_2$H$_4$O(Acetaldehyde)& -646.31 & -643.76 & -644.21 (2.10) & -641.73 (2.04) & -646.68 (-0.37) & -644.10 (-0.34) \\
C$_2$H$_4$O(Oxirane)& -620.55 & -618.25 & -618.38 (2.17) & -616.04 (2.21) & -620.92 (-0.36) & -618.49 (-0.24) \\
H$_2$O       & -223.57 & -223.26 & -222.99 (0.58) & -222.60 (0.66) & -223.90 (-0.34) & -223.51 (-0.25) \\
H$_2$O$_2$   & -256.30 & -255.78 & -255.10 (1.20) & -254.59 (1.19) & -256.48 (-0.18) & -255.98 (-0.20) \\
CO$_2$       & -366.44 & -364.51 & -364.59 (1.85) & -362.61 (1.90) & -366.76 (-0.32) & -364.70 (-0.19) \\
NH$_3$       & -291.30 & -290.65 & -290.54 (0.76) & -289.90 (0.75) & -291.59 (-0.29) & -290.93 (-0.28) \\
N$_2$H$_4$   & -427.91 & -426.72 & -426.26 (1.65) & -425.11 (1.61) & -428.22 (-0.31) & -427.04 (-0.32) \\
HCN          & -300.54 & -298.69 & -299.19 (1.35) & -297.42 (1.27) & -300.81 (-0.27) & -298.96 (-0.27) \\
C$_3$NH$_3$  & -725.79 & -721.71 & -723.18 (2.61) & -719.15 (2.56) & -726.37 (-0.58) & -722.16 (-0.45) \\
C$_2$NH$_5$  & -691.54 & -688.80 & -689.16 (2.38) & -686.48 (2.32) & -691.95 (-0.41) & -689.18 (-0.38) \\
CNH$_5$      & -563.31 & -561.65 & -561.78 (1.53) & -560.12 (1.53) & -563.74 (-0.43) & -562.03 (-0.38) \\
C$_2$N$_2$   & -476.28 & -472.61 & -473.92 (2.36) & -470.30 (2.31) & -476.78 (-0.50) & -472.99 (-0.38) \\
N$_2$O       & -259.68 & -258.10 & -257.83 (1.85) & -256.27 (1.82) & -259.84 (-0.16) & -258.24 (-0.14) \\
CHF$_3$      & -423.46 & -422.28 & -421.45 (2.01) & -420.28 (1.99) & -423.58 (-0.12) & -422.38 (-0.10) \\
C$_2$H$_3$F  & -542.26 & -539.87 & -540.42 (1.84) & -538.08 (1.79) & -542.61 (-0.35) & -540.18 (-0.31) \\
CH$_2$F$_2$  & -409.97 & -408.89 & -408.42 (1.55) & -407.30 (1.59) & -410.19 (-0.22) & -409.04 (-0.15) \\
NF$_3$       & -183.91 & -183.63 & -182.35 (1.56) & -182.02 (1.61) & -183.95 (-0.04) & -183.64 (-0.01) \\
C$_2$H$_3$OF & -667.47 & -664.80 & -665.00 (2.47) & -662.38 (2.42) & -667.86 (-0.39) & -665.15 (-0.35) \\
F$_2$O       & -78.95  & -78.79  & -77.96 (0.99) & -77.78 (1.01) & -78.98 (-0.03) & -78.81 (-0.02) \\
COF$_2$      & -386.78 & -385.16 & -384.65 (2.13) & -383.02 (2.14) & -386.92 (-0.15) & -385.25 (-0.09) \\
C$_3$H$_8$   & -967.94 & -964.66 & -965.64 (2.30) & -962.43 (2.23) & -968.52 (-0.58) & -965.19 (-0.53) \\
C$_6$H$_6$   & -1297.77
& -1291.50& -1293.79 (3.98) & -1286.95 (4.55) & -1299.02 (-0.62) & -1291.92 (-0.59) \\
H$_2$S       & -177.79 & -177.06 & -177.56 (0.23) & -176.87 (0.19) & -177.83 (-0.03) & -177.17 (-0.11) \\
CS$_2$       & -263.41 & -260.62 & -261.92 (1.49) & -259.40 (1.22) & -263.80 (-0.39) & -261.27 (-0.65) \\
CH$_4$S      & -456.48 & -454.85 & -455.61 (0.87) & -453.92 (0.93) & -456.91 (-0.43) & -455.25 (-0.40) \\
COS          & -315.45 & -313.28 & -314.02 (1.43) & -311.79 (1.49) & -315.97 (-0.52) & -313.68 (-0.40) \\
SO$_2$       & -243.33 & -241.90 & -241.49 (1.84) & -240.03 (1.87) & -243.69 (-0.36) & -242.25 (-0.35) \\
HClO         & -155.02 & -154.56 & -154.18 (0.84) & -153.68 (0.88) & -155.16 (-0.14) & -154.69 (-0.13) \\
CH$_3$Cl     & -377.24 & -375.96 & -376.33 (0.91) & -374.97 (0.99) & -377.61 (-0.37) & -376.25 (-0.29) \\
NOCl         & -181.15 & -180.24 & -179.82 (1.33) & -178.91 (1.33) & -181.21 (-0.06) & -180.29 (-0.05) \\
CH$_2$Cl$_2$ & -347.74 & -346.32 & -346.76 (0.98) & -345.12 (1.20) & -348.21 (-0.47) & -346.61 (-0.29) \\
C$_2$H$_3$Cl & -514.50 & -511.94 & -512.92 (1.58) & -510.36 (1.58) & -514.98 (-0.48) & -512.37 (-0.43) \\
C$_2$NH$_3$  & -589.88 & -586.93 & -587.85 (2.03) & -584.93 (2.00) & -590.30 (-0.42) & -587.26 (-0.33) \\
SiH$_4$      & -316.27 & -315.57 & -316.10 (0.17) & -315.38 (0.19) & -316.57 (-0.30) & -315.94 (-0.36) \\
PH$_3$       & -239.39 & -238.70 & -239.30 (0.09) & -238.55 (0.15) & -239.60 (-0.22) & -238.93 (-0.23) \\
BF$_3$       & -435.64 & -433.21 & -433.51 (2.12) & -431.27 (1.94) & -436.08 (-0.45) & -433.49 (-0.28) \\
\hline
ME          &   -      &   -      &  1.67  &  1.70  & -0.34 & -0.33 \\
MAE         &   -      &  -       & 1.67  & 1.70  & 0.34  & 0.33  \\
\hline
\end{tabularx}
\caption{The atomization energies of 50 molecules at the RPA level (including both the HF and the RPA correlation contributions) are reported. The second and third columns represent the AE and FC approximation atomization energies obtained in this work. The fourth and fifth columns show the all-electron (AE) and frozen-core (FC) atomization energies calculated with the aug-cc-pwCV5Z basis set. The sixth and seventh columns correspond to the extrapolated results obtained from the aug-cc-pwCVQZ and aug-cc-pwCV5Z data . The energies are given in unit of kcal/mol.}
\label{tab:rpa-bind}
\end{table*}

From Tab.~\ref{tab:rpa-bind}, it can be seen that the basis set error within the finite-basis SOS framework, with the largest aug-cc-pwCV5Z basis, is 1.7 kcal/mol (70 meV) for both all-electron and frozen-core RPA atomization energies. After extrapolating to the CBS limit, the mean error is reduced to about 0.34 kcal/mol. This indicates that the extrapolation approach can indeed significantly reduce the BSIEs. We also observe that the finite-basis results consistently underestimate the atomization energies, whereas the extrapolated results consistently overestimate them. We attribute this to the basis set superposition error (BSSE). In the calculations of Table~\ref{tab:rpa-bind}, neither the finite-basis nor the extrapolated results account for BSSE. To examine the influence of BSSE, we selected 10 molecules from Tab.~\ref{tab:rpa-bind} and investigated the effect of BSSE on the atomization energies. The results are presented in Table \ref{tab:bsse} in Appendix \ref{appendix:result-sup}. It can be seen that after the counterpoise (CP) correction, the tendency of the extrapolated results to overestimate the atomization energies is greatly reduced, while for certain molecules, an overshooting is observed. Overall, the CP correction still
changes the RPA atomization energies by 0.3 -- 0.5 kcal/mol even for the extrapolated results. Note that
the presented Delta-Sternheimer results do not suffer from BSSES, since the absolute RPA total energies are separately converged for the molecules and the individual atoms, and the results do not rely upon a
balanced description of the molecule and atoms.

We also checked the impact of the CP correction on finite-basis (aug-cc-pwCV5Z) results and found that the
influence is much larger here than on the extrapolated results, with the ME increasing from 0.95 kcal/mol to 3.22 kcal/mol, leading to a more pronounced underestimation of the atomization energy. Therefore, we arrive at the following guiding principle: for finite-basis calculations of RPA atomization energies, the results obtained without the CP correction are more reliable, whereas for results extrapolated to the basis-set limit, the CP correction can slightly improve accuracy.

\section{Summary}
\label{sec:summary}
In this work, by combining the high efficiency of atomic orbitals with the systematic nature of the FEM, we developed a FE Delta-Sternheimer approach, which can compute accurate RPA correlation energies for general polyatomic molecules at an affordable computational cost. We applied this method to determine the energy ordering of different configurations of water dimers and RPA atomization energies for 50 small molecules. These results are used to rigorously assess the finite basis set errors within the conventional SOS approach and the extrapolation method, from which the following guiding principles are obtained: For water clusters, to achieve meV-level accuracy of the energy difference, it is essential to use atomic basis sets at the QZ level that include diffuse functions. For the
RPA atomization energy of small molecules, if the results are obtained with finite AO basis sets, it is
advised that the BSSE correction is not performed. On the other hand, if an extrapolation to the CBS limit is performed, a BSSE correction is still helpful in rendering the results more accurate.

\section{Acknowledgment.}
We thank Prof. Volker Blum for inspiring discussions. We acknowledge the funding support by the National Key
Research and Development Program of China (Grant Nos.
2023YFA1507004, 2022YFA1403800 and 2025YFA1016601) and by the Strategic Priority
Research Program of Chinese Academy of Sciences
under Grant Nos. XDB0500201 and XDB0620203. This work was also supported by the National Natural Science Foundation of China (Grants Nos. 12134012, 12374067, 12188101 and 12331015). 
\appendix

\section{Finite element method and Mesh adaptive refinement}
\label{appendix:fem}
Here, we briefly outline the working principle of the FEM. The core of the FEM lies in constructing a FE space, and the first step in doing so is to partition the continuous domain. In one-, two‑, and three‑dimensional spaces, the domain is typically subdivided into line segments, triangles, and tetrahedral elements, respectively. Once the domain has been divided into a collection of elements, local basis functions are defined on each element. These local basis functions are nonzero only on the element itself and a small number of neighboring elements. They are usually chosen to be polynomial functions, with various possible forms and orders.  Taking the Laplace boundary‑value problem as an example to explain how FEM works, 
In a bounded domain $\Omega$, the continuous function $u$ satisfies,
\begin{equation}
-\nabla^2 u = f \quad  \text{inner}\, \, \text{of} \, \, \Omega,
\label{eq:Lap}
\end{equation}
\begin{equation}
u=0 \quad  \text{On}\, \,  \text{the}\, \, \text{boundary}\,\,  \partial \Omega. 
\end{equation}
Suppose we have constructed the FE space $V_h$, 
which contains the collection of FE basis functions $\{\varphi_j\}$. 
In the FE space $V_h$, we let $u_h$ denote the approximate solution of 
the partial differential Eq. \ref{eq:Lap},
\begin{equation}
u_h = \sum_j^{N_h} c_j \varphi_j. 
\label{eq:fem_sol}
\end{equation}
Here, $N_h$ is the total number of FE basis. The problem is reduced to solving for the expansion coefficients ${c_j}$. By substituting Eq.~\ref{eq:fem_sol} into Eq.~\ref{eq:Lap}, then multiplying on the left by the FE basis function $\varphi_i$ and integrating over real space, we obtain,
\begin{equation}
\label{eq:bd_eq}
\sum_j^{N_h} a(i,j)c_j = f_i.
\end{equation}
Here, $f_i=\int_\Omega f({\bf r})\varphi_i({\bf r})d\bf{r}$ and $a(i,j)$ represents 
\begin{equation}
\begin{aligned}
a(i,j) &= \int_\Omega -\varphi_i({\bf{r}}) \nabla^2 \varphi_j({\bf r})d\bf{r}\\
       &=\int_\Omega \nabla \varphi_i({\bf{r}}) . \nabla \varphi_j({\bf r})  d\bf{r} \, .
\end{aligned}
\label{eq:aij}
\end{equation}
In Eq.~\ref{eq:aij}, we apply the integration‑by‑parts formula and use the fact that the FE basis functions vanish on the boundary. 
From Eq.~\ref{eq:bd_eq}, we see that in the FE space, the partial differential equation is discretized into a system of linear equations. Moreover, owing to the locality of the FE basis functions, the working matrix (commonly called the stiffness matrix) is sparse. This sparsity provides tremendous advantages for practical computations.

In the FEM, adaptive mesh refinement plays a crucial role in enhancing both computational efficiency and solution accuracy. Because of the complexity of real-world problems, it is often impossible to determine the optimal mesh \textit{a priori}. Adaptive refinement uses \textit{a posteriori} error estimates to automatically identify regions requiring finer resolution, thus minimizing computational cost while maintaining accuracy. 

The core idea of adaptive refinement is to locally enrich the mesh in areas where the estimated error is large, while keeping a coarser mesh in regions where the error is small. This strategy not only effectively improves accuracy, but also avoids the dramatic increase in computational workload associated with global refinement. By iterating the adaptive refinement process several times, the mesh distribution can be progressively optimized, concentrating computational effort precisely where it is most needed.

In the adaptive refinement process, a posteriori error estimator is used to generate an error indicator for each mesh element. For the boundary-value problem Eq.~\ref{eq:Lap}, 
the error indicator $\eta_K$ on the element $K$ is defined as follows,
\begin{equation}
\eta_{K}^{2}\left(u_{h}, f\right)=h_{K}^{2}\left\|f+\nabla^2  u_{h}\right\|_{0, K}^{2}+\sum_{e \in \mathcal{E}_{K}} h_{e}\left\|\left[\frac{\partial u_{h}}{\partial \mathbf{n}}\right]\right\|_{0, e}^{2},
\label{eq:error_indicator}
\end{equation}
where, $h_K$ denotes the diameter of element $K$, and 
$h_e$ denotes the diameter of a face (3 dimensional) or an edge (2 dimensional) of element $K$. 
The quantity $\frac{\partial u_h}{\partial \bf{n}}$ 
represents the derivative of the FE solution $u_h$ 
in the outward normal direction on the face (3 dimensional) or edge (2 dimensional) $e$. $\mathcal{E}_{K}$ denotes the set of all faces (3 dimensional) or edges (2 dimensional) 
of element $K$.

Accordingly, the global a posteriori error estimate can be defined as follows,
\begin{equation}
\eta_{h}^{2}\left(u_{h}, f\right)=\sum_{K\in \mathcal{T}_h} \eta_{K}^{2}\left(u_{h}, f\right),
\end{equation}
where $\mathcal{T}_h$ denotes the FE mesh.

We adopt the D\"{o}rfler marking strategy to select the elements requiring 
local refinement based on the error indicator Eq.~\ref{eq:error_indicator} \cite{Dorfler1996},\\
(1). Sorting: Arrange all elements in descending order of their error indicators, 
i.e. $\eta_{K_1} \ge \eta_{K_2} \ge \cdots \ge \eta_{K_N}$.\\
(2). Select the smallest integer  M such that the sum of the errors over the first M elements exceeds a given threshold 
$\theta \in \left( 0,1\right)$ of the total error,
\begin{equation}
\sum_{i}^M \eta_{K_i}^{2} \ge \theta \, \eta_{h}^{2}\left(u_{h}, f\right) \, .
\end{equation}

In Sec.~I.B, we discussed the  adaptive mesh refinement approach for FE mesh based on posteriori error. It should be noted that the  posteriori error estimator is defined with respect to a particular differential equation. In our work, for a given molecular configuration, we need to solve the Sternheimer equation for different occupied‐state orbitals, different spatial distributions of the perturbation, and different frequency points. If we  perform adaptive refinement separately for each of these equations, we would obtain extremely similar meshes, resulting in a substantial waste of computational resources.

Here, we choose instead to solve the Sternheimer equation 
for all occupied states in the static case ($\omega = 0$) 
under a single representative perturbation. 
We then define the total posteriori error on each element 
as the sum of the local error estimators for each occupied‐state orbital. 
In this way, we generate one FE mesh tailored to the molecule. 
The choice of representative perturbation is also very 
simple and intuitive. We choose,
\begin{equation}
V^{(1)}(\bf{r}) = \alpha V_{ext}(\bf{r}), 
\end{equation}
where $\alpha$ is set to 0.1. Such a choice is physically 
well motivated, since the external potential naturally 
encodes the positions and atomic numbers of the constituent atoms. 
The corresponding FE mesh generated from this perturbation  
exhibits a desirable feature: it is refined in the vicinity 
of atomic nuclei and becomes gradually coarser in regions farther away.

\section{Converge behavior with respect to cubic box size and basis set for describing the occupied manifold}
\label{appendix:converge-test}
In this section, we first examine the effect of the choice of the simulation box on the calculations. The box size determines the real space where the molecule resides: If the box is too small, the wavefunction does not decay to zero at the boundaries; if the box is too large, computational resources are wasted. Therefore, we first tested the RPA correlation energy of the acetylene molecule using different box sizes. The basis set employed was aug-cc-pwCV5Z, and the pre-optimized auxiliary basis set aug-cc-pwCV5Z-RI-FITTING was used. Box sizes of 30, 40, and 50 Bohr were considered. All other parameters were kept the same, and the resulting RPA correlation energies are summarized as follows:

\begin{table}[htbp]
\centering
\footnotesize
\caption{Convergence test of RPA correlation energy with respect to box size. The energy unit is eV.}
\label{tab:Rmax}
\begin{tabular}{cccc}
\hline
Molecule & 30 & 40 & 50 \\
\hline
C$_2$H$_2$ & -20.91153 & -20.91151 & -20.91149 \\
PH$_3$     & -27.61443 & -27.61470   & -27.61479   \\
\hline
\end{tabular}
\end{table}

From Tab.~\ref{tab:Rmax}, it can be seen that, as the size of the cubic box increases from 30 Bohr to 50 Bohr, the change in the correlation energy is less than 0.1 meV. This indicates that in the present calculations the box size does not need to be particularly large. It should be noted that in Ref.~\cite{peng2024textit} we also discussed the effect of the choice of infinity in prolate spheroidal coordinates on the correlation energy, where we found that setting infinity at 40 Bohr still leads to an absolute correlation energy error of several meV. In contrast, the convergence with respect to infinity in the present work is much faster. The difference arises from the fact that the present calculations are performed at the RI-RPA level. In our approach, the auxiliary basis functions localized within a few Bohr around the atomic centers are regarded as perturbations to the system, and the resulting first-order density is subsequently projected back onto the auxiliary basis functions. Therefore, the computational framework employed here is insensitive to the choice of infinity. 

Next, we examine the completeness of the occupied-state manifold described by atomic orbitals. In this work, only the occupied states obtained from atomic orbital basis-set calculations are required. The basis set affects two parts of the total RPA energy specifically: (1) the RPA correlation energy, where the occupied-state energies and wavefunctions used in solving the Sternheimer equation in the FE space depend on the chosen basis set; and (2) the non-self-consistent HF energy, where the single particle energies, the Hartree energy and exact-exchange energy only depend on the occupied manifold. In this work, we primarily employ the aug-cc-pwCV5Z basis set, as we have tested its accuracy in terms of both the RPA correlation energy and the non-self-consistent HF energy.

To examine the convergence of the RPA correlation energy with respect to the single-particle basis set describing the occupied-state manifold, we performed calculations starting from aug-cc-pwCVTZ, aug-cc-pwCVQZ, and aug-cc-pwCV5Z, and analyzed the dependence of the RPA correlation energy on the basis set. To eliminate the influence of the auxiliary basis, the same pre-optimized aug-cc-pwCV5Z-RI-FITTING auxiliary basis set was used for all single-particle basis sets. The box size was set to 30 Bohr, and identical FE grids were used for all calculations, with 32 Gauss–Legendre frequency points. The results are summarized in Tab.~\ref{tab:converge_occ}. It can be observed that the difference between the QZ and 5Z calculations is less than 1 meV. Even when using only the TZ basis for the DFT calculation, the impact on the RPA correlation energy is merely 1–2 meV. Therefore, the RPA correlation energy converges rapidly with respect to the atomic orbital basis set describing the occupied-state manifold.

With respect to the basis set convergence of the non-self-consistent HF energy, detailed tests are provided in Table~\ref{tab:exx-bind} in the following section. The results demonstrate that, at the level of atomization energies, the non-self-consistent HF contribution obtained with the aug-cc-pwCV5Z basis set converges to the meV level.

\begin{table}[htbp]
\centering
\caption{Converge behavior with respect to the basis set used for describing the occupied manifold. The energy unit is eV.}
\label{tab:converge_occ}
\footnotesize
\begin{tabular}{cccc}
\hline
Molecule & TZ & QZ & 5Z \\
\hline
C$_2$H$_2$ & -20.90956 & -20.91073 & -20.91153 \\
PH$_3$     & -27.61540 & -27.61416 & -27.61443 \\
C$_3$H$_6$ & -34.64236 & -34.64270 & -34.64356 \\
H$_2$O     & -15.76272 & -15.76281 & -15.76321 \\
\hline
\end{tabular}
\end{table}

\section{Details of the computation of the residual function
$D_a$}
\label{appendix:Da}
To evaluate $\langle D_a | \psi_{i,\text{out}}^{(1)}(i\omega)\rangle$ in Eq.\ref{eq:1-order-total}, we need to compute the value of the residual function $D_a$ at each FE quadrature point. Suppose that a DFT calculation is performed using an AO basis set $\{ \phi_{j,AO}   \}$, yielding a set of eigenstates $\{ \psi_{j,AO} \} $ .  
For an arbitrary unoccupied state $\{ \psi_{a,AO} \} $, one obtains from Eq.\ref{eq:H0_on_psia} that,
\begin{equation}
\begin{aligned}
D_a
&= \left(-\frac{1}{2}\nabla^2 + V_{\text{eff}} - \epsilon_{a,\text{AO}}\right)\psi_{a,\text{AO}} \\
&= \sum_{j=1}^{N_b} c_{aj}
\left(-\frac{1}{2}\nabla^2 + V_{\text{eff}} - \epsilon_{a,\text{AO}}\right)\phi_{j,\text{AO}} .
\end{aligned}
\label{eq:Da}
\end{equation}
Here, $N_\text{b}$ represents the number of AO basis functions. $c_{aj}$ represents the KS eigenvector. 
It is straightforward to see from Eq.~\ref{eq:Da} that the evaluation of the residual function requires, at each FE quadrature point, the values of the AO basis functions, the effective potential, and the action of the kinetic-energy operator on the basis functions.
The values of the effective potential and the AO basis functions at an arbitrary real-space point can be obtained accurately using one-dimensional interpolation, as discussed in our previous work \cite{peng2024textit}.
Moreover, the action of the kinetic-energy operator on an AO basis function preserves the form of a radial function multiplied by a spherical harmonic. It can therefore be evaluated accurately on a one-dimensional logarithmic radial grid, and FHI-aims can directly output the resulting radial functions. Consequently, the values at arbitrary spatial points can again be obtained using one-dimensional interpolation.
Finally, the residual function $D_a$ at each FE quadrature point is obtained using Eq.\ref{eq:Da}.

\section{RI error test}
\label{appendix:convergence_RI_error}
In Section~\ref{sec:convergence_benchmark},  we tested the convergence of the RPA correlation energy with respect to the FE grid. The results show that when the  NOF reaches approximately 300,000, the RPA correlation energy can converge to the meV level. Consequently, in the RPA calculations, the SPBE which represents the most significant part of the basis set error, has been almost completely eliminated (the SPBE associated with the description of the occupied-state manifold will be discussed in the next section). However, within the RI framework, one must also ensure the completeness of the ABS. Although Ref.~\cite{peng2023basis} shows that the incompleteness error of the ABS is much smaller than the SPBE, it is still visible in practical calculations. In Ref.~\cite{peng2023basis} and Ref.~\cite{peng2024textit}, we completely eliminated the RI approximation by employing iterative diagonalization to obtain  basis-error-free results. While iterative diagonalization could likewise eliminate the RI errors here, it is prohibitively expensive. For the mono- and diatomic systems treated in Refs.~\cite{peng2023basis} and \cite{peng2024textit}, full convergence of the correlation energy is achievable, but it becomes exceedingly difficult for the general polyatomic molecules considered in this work. Accordingly, we compute RPA correlation energies within the RI framework in this work  and perform extensive tests to quantify the RI error. 

Here we examine the RI-RPA atomization energies of a series of diatomic molecules in the prolate spheroidal coordinate system of Ref.~\cite{peng2024textit}, and compare them to the basis-error-free results reported in Ref.~\cite{peng2024textit}, thereby estimating the RI error in the calculated atomization energy. We stress that, since solving the Sternheimer equation in prolate spherical  coordinates eliminates SPBE, the differences in atomization energies obtained via the RI approach compared to the reference values of Ref.~\cite{peng2024textit} arise entirely from the ABS. Therefore, these differences can be taken as a measure of the RI error.  The tests include comparisons of errors from (i) ABS generated on-the-fly from single-particle basis sets versus optimal RI-fitting ABS, (ii) ABSs generated on-the-fly from different types of correlation-consistent basis sets, and (iii) ABSs with varying maximum angular momenta.
We tested the effects of ABSs generated on the fly using the aug-cc-pwCV5Z and NAO-VCC-5Z basis sets on the atomization energies of a series of diatomic molecules. Two cases for the maximum angular momentum of the auxiliary basis functions, 5 and 9, were considered. In addition, we tested the pre-optimized ABS aug-cc-pwCV5Z-RI-FITTING. The results are presented in Tables Tab.~\ref{tab:combined-RI}, while Tab.~\ref{tab:abs-size} lists the sizes of the different ABSs used in this test for the purpose of evaluating computational efficiency.

\begin{table*}[htbp]
\centering
\footnotesize
\renewcommand{\arraystretch}{1.5} 
\begin{tabular}{|>{\centering\arraybackslash}m{1.2cm}|>{\centering\arraybackslash}m{1.0cm}|
                >{\centering\arraybackslash}m{2.0cm}|>{\centering\arraybackslash}m{2.0cm}|
                >{\centering\arraybackslash}m{2.0cm}|>{\centering\arraybackslash}m{2.0cm}|}
\hline
\multirow{2}{*}{\textbf{ATOM}} & \multirow{2}{*}{\textbf{OPT}} 
  & \multicolumn{2}{c|}{\textbf{aug-cc-pwCV5Z}} & \multicolumn{2}{c|}{\textbf{NAO-VCC-5Z}} \\
\cline{3-6}
 & & $L_{\max}=5$ & $L_{\max}=9$ & $L_{\max}=5$ & $L_{\max}=9$ \\
\hline
H  & 189 & 412  & 863  & 224 & 310 \\
C  & 323 & 669  & 1311 & 362 & 589 \\
N  & 323 & 664  & 1306 & 366 & 593 \\
O  & 324 & 675  & 1332 & 352 & 579 \\
F  & 324 & 665  & 1322 & 374 & 629 \\
P  & 411 & 875  & 1805 & 484 & 769 \\
Cl & 411 & 879  & 1796 & 505 & 807 \\
\hline
\end{tabular}
\caption{Basis-set size of different ABS for different elements.}
\label{tab:abs-size}
\end{table*}

\begingroup
\setlength{\extrarowheight}{3pt}
\renewcommand{\arraystretch}{1.25}

\begin{table*}[!htbp]
\centering
\caption{RI-RPA atomization energy for several diatomic molecules. OTF-Gaussian denotes the auxiliary basis generated on-the-fly (OTF) with the aug-cc-pwCV5Z basis set, while OTF-NAO denotes the auxiliary basis generated on-the-fly with the NAO-VCC-5Z basis set. Lmax represents the maximum angular momentum of the OTF auxiliary basis. OPT refers to the pre-optimized auxiliary basis aug-cc-pwCV5Z-RIFIT. $\Delta$ indicates the deviation from the reference, with the reference taken from Ref.~\cite{peng2024textit}.}
\label{tab:combined-RI}
\begin{tabular}{>{\centering\arraybackslash}c|cccc|cccc|>{\centering\arraybackslash}p{1.2cm}>{\centering\arraybackslash}p{1.2cm}|>{\centering\arraybackslash}p{1.8cm}}
\hline
\multirow{2}{*}{\textbf{MO}} 
  & \multicolumn{4}{c|}{\textbf{OTF-Gaussian}} 
  & \multicolumn{4}{c|}{\textbf{OTF-NAO}} 
  & \multicolumn{2}{c|}{\textbf{OPT}} 
  & \multirow{2}{*}{\textbf{Reference}} \\
\cline{2-11} 
 & \textbf{$L_{\text{max}}$=5} & \textbf{$L_{\text{max}}$=9} & \textbf{$\Delta$(5)} & \textbf{$\Delta$(9)}
 & \textbf{$L_{\text{max}}$=5} & \textbf{$L_{\text{max}}$=9} & \textbf{$\Delta$(5)} & \textbf{$\Delta$(9)}
 & \textbf{RI} & \textbf{$\Delta$} & \\  
\hline
N$_2$  & -224.16 & -224.36 &  0.24 & 0.04 & -223.91 & -224.14 & 0.49 & 0.26 & -223.97 & 0.43 & -224.40 \\
O$_2$  & -113.65 & -113.68 &  0.14 & 0.11 & -113.37 & -113.48 & 0.42 & 0.31 & -113.49 & 0.30 & -113.79 \\
P$_2$  & -116.80 & -117.14 &  0.39 & 0.05 & -116.63 & -117.16 & 0.56 & 0.03 & -116.65 & 0.54 & -117.19 \\
CO   & -245.48 & -245.56 &  0.13 & 0.06 & -245.36 & -245.51 & 0.25 & 0.10 & -245.27 & 0.34 & -245.61 \\
H$_2$  & -108.73 & -108.74 & -0.01 & -0.02 & -108.71 & -108.71 & 0.01 & 0.01 & -108.68 & 0.04 & -108.72 \\
Cl$_2$ &  -50.11 &  -50.18 &  0.07 & 0.00 &  -49.54 &  -49.79 & 0.64 & 0.39 &  -49.86 & 0.32 & -50.18 \\
HF   & -132.93 & -132.76 & -0.16 & 0.01 & -132.88 & -132.78 & -0.11 & -0.01 & -132.66 & 0.11 & -132.77 \\
F$_2$  &  -30.54 &  -30.61 &  0.07 & 0.00 &  -30.31 &  -30.44 & 0.30 & 0.17 &  -30.48 & 0.13 & -30.61 \\
\hline
ME     & -- & -- & 0.11 & 0.03 & -- & -- & 0.32 & 0.16 & -- & 0.28 & -- \\
MAE    & -- & -- & 0.16 & 0.04 & -- & -- & 0.35 & 0.16 & -- & 0.28 & -- \\
\hline\hline
\end{tabular}
\end{table*}

\endgroup
It can be seen that the ABS generated on-the-fly using aug-cc-pwCV5Z yields the smallest error. When the maximum angular momentum is set to 9, the results deviate from the reference by only 0.04 kcal/mol (about 2 meV). The cost, however, is 1,000–2,000 auxiliary basis functions per atom. We also note that increasing the maximum angular momentum of the ABS is an effective way to reduce the RI error, although this also leads to a significant increase in the basis size. Another point worth mentioning is the efficiency of the pre-optimized auxiliary basis set. With only 300–400 auxiliary basis functions per atom, the RI error can be reduced to 0.28 kcal/mol (about 12 meV), corresponding to approximately 10 meV. Therefore, different auxiliary basis sets can be chosen depending on the desired computational precision. In this work, solving the Delta-Sternheimer equation within the FE framework allows us to almost completely eliminate the single-particle basis set error (see Sec. III A). To further minimize the RI error, we  employ the aug-cc-pwCV5Z basis set and set the maximum angular momentum of the auxiliary basis functions to 9 when we calculate the atomization energy of molecules.

\section{Further results on the energy hierarchy of water dimers and molecular atomization energies }
\label{appendix:result-sup}
\subsection{Water dimer}
In the main text, we presented a schematic plot of the total energies of twenty different water dimer configurations obtained with the aug-cc-pwCVXZ basis sets. In Fig.~\ref{fig:water-dimer-sup} below, we show the corresponding results obtained with the cc-pVXZ and NAO-VCC-nZ basis sets.
\begin{figure*}[htbp]
   \centering
    \includegraphics[scale=0.4]{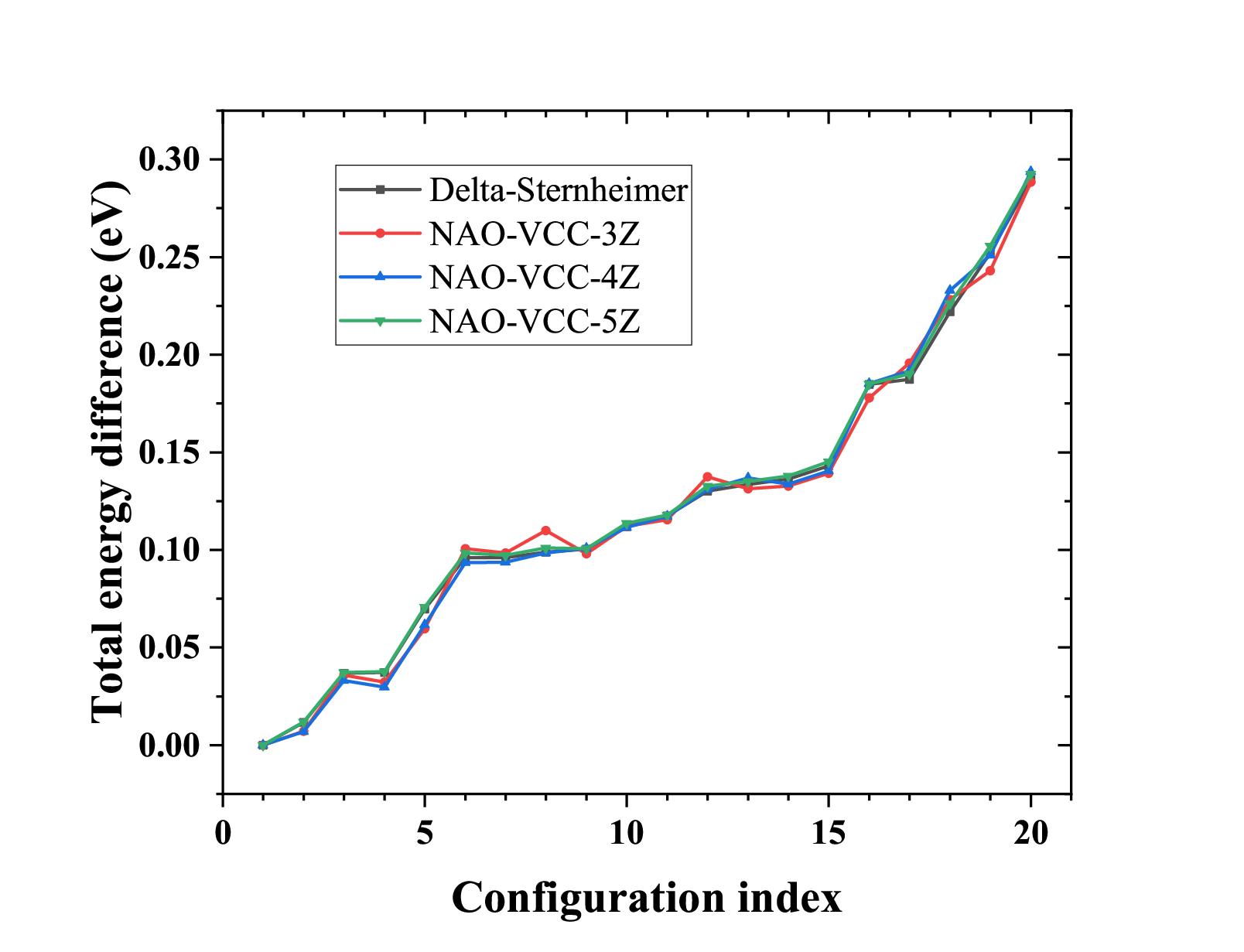}
    \\
    \includegraphics[scale=0.4]{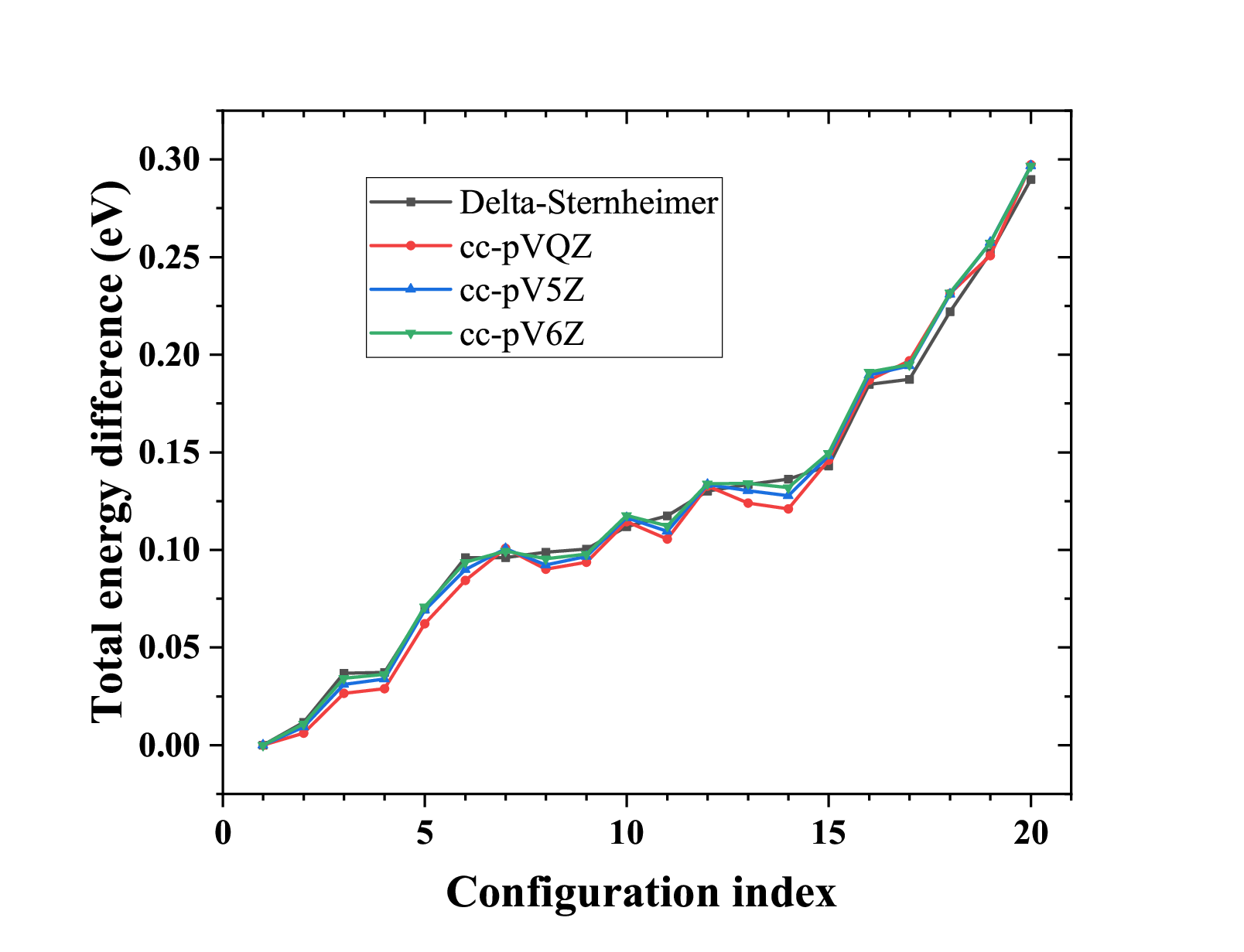}
    \caption{ The black lines in the upper and lower panels, as well as the black line in Fig. \ref{fig:water-dimer}, correspond to the same data set and represent the RPA total energies obtained with the FE Delta-Sternheimer method developed in this work. The results compared against the black line are obtained using correlation-consistent basis sets of different sizes and from different basis-set families.  } 
    \label{fig:water-dimer-sup}
\end{figure*}

\subsection{Atomization energy }
 To evaluate the basis set error of the HF part to the atomization energies, we performed calculations using aug-cc-pwCVQZ and aug-cc-pwCV5Z basis sets and extrapolated the results to the CBS limit using a standard formula Eq.~\ref{eq:exx-extra},
\begin{equation}
E_{\rm HF}^{XY} = 
\frac{e^{-a \sqrt{Y}}\,E_{\rm HF}^{X} - e^{-a \sqrt{X}}\,E_{\rm HF}^{Y}}
     {e^{-a\sqrt{Y}} - e^{-a\sqrt{X}} }.
     \label{eq:exx-extra}
\end{equation}
\begin{equation}
\label{eq:rpa-extra}
E_{\rm corr}^{XY} = \frac{X^3 E_{\rm corr}^{X} - Y^3 E_{\rm corr}^{Y}}{X^3 - Y^3}.
\end{equation}

In addition, calculations with the built-in NAO ``\textit{Tier4+}" basis sets of FHI-aims were also performed for reference. These results are summarized in Tab.~\ref{tab:exx-bind}.

\newcolumntype{Y}{>{\centering\arraybackslash}p{3.2cm}}
\newcolumntype{C}{>{\centering\arraybackslash}X} 

\begin{table*}[htbp]
\centering
\scriptsize
\renewcommand{\arraystretch}{1.2}
\begin{tabularx}{\textwidth}{|Y|C|C|C|C|C|C|}
\hline
Molecule & QZ & 5Z & Extrapolation & Tier & $\Delta$-Extrapolation & $\Delta$-Tier \\
\hline
CH$_4$ & -327.37 & -327.50 & -327.54 & -327.49 & 0.04 & 0.01 \\
C$_2$H$_2$ & -291.07 & -291.20 & -291.24 & -291.21 & 0.04 & 0.01 \\
C$_2$H$_4$ & -425.16 & -425.32 & -425.36 & -425.33 & 0.05 & 0.01 \\
C$_3$H$_6$(Propene) & -649.73 & -649.95 & -650.02 & -649.96 & 0.06 & 0.01 \\
C$_3$H$_6$(Cyclopropane) & -641.50 & -641.72 & -641.78 & -641.72 & 0.06 & 0.00 \\
C$_3$H$_4$(Allene) & -516.37 & -516.55 & -516.61 & -516.57 & 0.05 & 0.02 \\
C$_3$H$_4$(Propyne) & -519.41 & -519.60 & -519.65 & -519.62 & 0.05 & 0.02 \\
C$_3$H$_4$(Cyclopropene) & -493.00 & -493.18 & -493.24 & -493.21 & 0.05 & 0.02 \\
C$_2$H$_6$ & -548.57 & -548.77 & -548.83 & -548.76 & 0.06 & 0.01 \\
CH$_2$O & -251.67 & -251.71 & -251.72 & -251.71 & 0.01 & 0.00 \\
CH$_3$OH & -365.21 & -365.31 & -365.34 & -365.30 & 0.03 & 0.02 \\
CH$_2$O$_2$ & -319.40 & -319.40 & -319.40 & -319.37 & 0.00 & 0.03 \\
C$_2$H$_2$O$_2$ & -407.97 & -407.98 & -407.98 & -407.97 & 0.00 & 0.02 \\
C$_2$H$_6$O & -590.32 & -590.47 & -590.51 & -590.44 & 0.04 & 0.03 \\
C$_2$H$_4$O(Acetaldehyde) & -482.09 & -482.19 & -482.22 & -482.17 & 0.03 & 0.02 \\
C$_2$H$_4$O(Oxirane) & -452.68 & -452.77 & -452.80 & -452.78 & 0.03 & 0.01 \\
H$_2$O & -154.87 & -154.87 & -154.88 & -154.88 & 0.00 & 0.00 \\
H$_2$O$_2$ & -129.37 & -129.36 & -129.35 & -129.31 & 0.00 & 0.05 \\
CO$_2$ & -234.03 & -233.97 & -233.95 & -233.94 & 0.02 & 0.03 \\
NH$_3$ & -199.64 & -199.71 & -199.73 & -199.70 & 0.02 & 0.00 \\
N$_2$H$_4$ & -263.71 & -263.81 & -263.84 & -263.79 & 0.03 & 0.02 \\
HCN & -195.13 & -195.20 & -195.21 & -195.19 & 0.02 & 0.01 \\
C$_3$NH$_3$ & -525.11 & -525.26 & -525.31 & -525.27 & 0.04 & 0.01 \\
C$_2$NH$_5$ & -501.87 & -502.03 & -502.08 & -502.03 & 0.05 & 0.01 \\
CNH$_5$ & -413.16 & -413.31 & -413.35 & -413.29 & 0.04 & 0.02 \\
C$_2$N$_2$ & -280.52 & -280.57 & -280.59 & -280.57 & 0.02 & 0.01 \\
N$_2$O & -72.73 & -72.74 & -72.74 & -72.68 & 0.00 & 0.06 \\
CHF$_3$ & -299.01 & -298.93 & -298.90 & -298.94 & 0.02 & 0.02 \\
C$_2$H$_3$F & -410.40 & -410.49 & -410.51 & -410.50 & 0.03 & 0.01 \\
CH$_2$F$_2$ & -300.42 & -300.43 & -300.43 & -300.42 & 0.00 & 0.01 \\
NF$_3$ & -14.44 & -14.34 & -14.31 & -14.30 & 0.03 & 0.04 \\
C$_2$H$_3$OF & -486.04 & -486.07 & -486.08 & -486.05 & 0.01 & 0.02 \\
F$_2$O & 55.73 & 55.75 & 55.76 & 55.80 & 0.01 & 0.04 \\
COF$_2$ & -249.56 & -249.50 & -249.48 & -249.48 & 0.02 & 0.02 \\
C$_3$H$_8$ & -770.57 & -770.82 & -770.90 & -770.81 & 0.08 & 0.01 \\
C$_6$H$_6$ & -1015.40 & -1015.67 & -1015.75 & -1015.73 & 0.08 & 0.06 \\
H$_2$S & -128.41 & -128.45 & -128.46 & -128.48 & 0.01 & 0.03 \\
CS$_2$ & -156.04 & -156.10 & -156.11 & -156.17 & 0.02 & 0.07 \\
CH$_4$S & -347.43 & -347.57 & -347.61 & -347.59 & 0.04 & 0.03 \\
COS & -196.31 & -196.33 & -196.34 & -196.36 & 0.01 & 0.03 \\
SO$_2$ & -92.76 & -92.85 & -92.88 & -92.75 & 0.03 & 0.10 \\
HClO & -73.34 & -73.35 & -73.36 & -73.33 & 0.00 & 0.03 \\
CH$_3$Cl & -294.11 & -294.23 & -294.26 & -294.22 & 0.03 & 0.01 \\
NOCl & -27.15 & -27.11 & -27.09 & -27.06 & 0.01 & 0.05 \\
CH$_2$Cl$_2$ & -256.93 & -257.00 & -257.01 & -257.04 & 0.02 & 0.04 \\
C$_2$H$_3$Cl & -392.62 & -392.75 & -392.78 & -392.77 & 0.04 & 0.02 \\
C$_2$NH$_3$ & -427.34 & -427.47 & -427.51 & -427.46 & 0.04 & 0.01 \\
SiH$_4$ & -252.51 & -252.62 & -252.65 & -252.66 & 0.03 & 0.04 \\
PH$_3$ & -168.38 & -168.45 & -168.47 & -168.47 & 0.02 & 0.02 \\
BF$_3$ & -351.41 & -351.26 & -351.21 & -351.31 & 0.04 & 0.06 \\
\hline
MAE & - &  -& - & - & 0.03 & 0.02 \\
\hline
\end{tabularx}
\caption{Atomization energy of non-SCF HF part for 50 molecules. 
The energy unit is kcal/mol. The second and third columns present results obtained with the aug-cc-pwCVQZ and aug-cc-pwCV5Z basis sets, respectively. The fourth column shows the CBS-extrapolated values obtained from the second and third columns using the two-point extrapolation formula (Eq.~\ref{eq:exx-extra}). The fifth column provides the results from the largest available ``\textit{tier}" basis set in FHI-aims (i.e., tier 4 plus additional basis functions). In the sixth and seventh columns, we report the absolute deviations of the aug-cc-pwCV5Z results from the extrapolated values and the NAO basis set results, respectively.}
\label{tab:exx-bind}
\end{table*}

To examine the influence of BSSE, we selected 10 molecules from Tab.~\ref{tab:rpa-bind} and consider the counterpoise (CP) correction to the BSSEs. The results are presented in Tab.~\ref{tab:bsse}.

\begin{table*}[htbp]
\centering
\setlength{\tabcolsep}{6pt}
\renewcommand{\arraystretch}{1.2}
\begin{tabular}{|c|c|c|c|c|c|}
\hline
Molecule & 5Z (CP) & 5Z & Extra (CP) & Extra & This work \\
\hline
H$_2$O    & -221.80 (1.77) & -222.99 (0.58) & -223.58 (-0.01) & -223.90 (-0.33) & -223.57 \\
NH$_3$    & -289.17 (2.13) & -290.54 (0.76) & -291.29 (0.00) & -291.59 (-0.29) & -291.30 \\
CO$_2$    & -362.38 (4.06) & -364.59 (1.85) & -366.49 (-0.05) & -366.76 (-0.32) & -366.44 \\
CH$_2$O   & -353.93 (3.03) & -355.69 (1.27) & -356.89 (0.07) & -357.17 (-0.21) & -356.97 \\
C$_2$H$_6$   & -681.87 (4.24) & -684.54 (1.57) & -686.04 (0.08) & -686.51 (-0.40) & -686.11 \\
C$_3$H$_6$-2 & -810.44 (5.88) & -815.37 (0.95) & -816.20 (0.12) & -816.75 (-0.43) & -816.32 \\
C$_2$NH$_3$  & -584.71 (5.17) & -587.85 (2.02) & -589.83 (0.05) & -590.30 (-0.42) & -589.88 \\
SiH$_4$   & -314.21 (2.06) & -316.10 (0.17) & -315.98 (0.29) & -316.57 (-0.30) & -316.27 \\
PH$_3$    & -237.47 (1.92) & -239.30 (0.08) & -239.09 (0.29) & -239.60 (-0.22) & -239.39 \\
H$_2$S    & -175.90 (1.89) & -177.56 (0.23) & -177.41 (0.38) & -177.83 (-0.03) & -177.79 \\
\hline
ME     & 3.22 & 0.95 & 0.12 & -0.30 & - \\
MAE    & 3.22 & 0.95 & 0.14 & 0.30 &- \\
\hline
\end{tabular}
\caption{Counterpoise (CP) correction to the BSSEs for 10 selected molecules. The second and third columns report the atomization energies obtained using aug-cc-pwCV5Z basis set with and without CP correction, respectively. The fourth and fifth columns give the corresponding extrapolated CBS(Q,5) results. The final column presents the reference values obtained in this work, with the numbers in parentheses indicating the deviations from the reference results. All values are given in units of kcal/mol. }
\label{tab:bsse}
\end{table*}

\bibliography{RPA_bib}

\end{document}